\documentclass[
preprintnumbers,%
prc,%
10pt,%
final,%
notitlepage,%
oneside,%
twocolumn,%
nobibnotes,%
nofootinbib,% In the current version of REVTeX, when this option is on,
superscriptaddress,%
floatfix,%
showkeys,%
showpacs]%
{revtex4}
\usepackage{color} %{\color{red} }
\usepackage{amsfonts}
\usepackage{amsbsy}
\usepackage{mathrsfs}
\usepackage{graphicx}
\def\lsim{\mathrel{\rlap{
\lower4pt\hbox{\hskip-3pt$\sim$}}
    \raise1pt\hbox{$<$}}}     %less than approx. symbol
\def\gsim{\mathrel{\rlap{
\lower4pt\hbox{\hskip-3pt$\sim$}}
    \raise1pt\hbox{$>$}}}     %greater than or approx. symbol
\def\scr#1{\mbox{\scriptsize #1}}

\begin{document}
%\title{Three-fluid hydrodynamics based event simulation for collisions \\ at NICA and FAIR energies}%
\title{Event simulation based on three-fluid hydrodynamics for collisions at energies available at the Dubna Nuclotron-based Ion Collider Facility and at the \\Facility for Antiproton and Ion Research in Darmstadt}%
\author{P. Batyuk}\thanks{e-mail:  pavel.batyuk@jinr.ru}
\affiliation{Veksler and Baldin Laboratory of High Energy Physics, JINR Dubna, 141980 Dubna, Russia}
\author{D. Blaschke}\thanks{e-mail:  blaschke@ift.uni.wroc.pl}
\affiliation{Institute of Theoretical Physics, University of Wroclaw, 50-204 Wroclaw, Poland}
\affiliation{Bogoliubov Laboratory of Theoretical Physics, JINR Dubna, 141980 Dubna, Russia}
\affiliation{National Research Nuclear University "MEPhI" (Moscow Engineering
Physics Institute), 115409 Moscow, Russia} 
\author{M. Bleicher}\thanks{e-mail:  bleicher@th.physik.uni-frankfurt.de}
\affiliation{Frankfurt Institute for Advanced Studies (FIAS),
%Room 2|401
Science Campus Riedberg, Ruth-Moufang-Strasse 1, 60438 Frankfurt am Main, Germany}
\affiliation{Institut f\"ur Theoretische Physik, Goethe Universit\"at, Max-von-Laue-Strasse 1, 60438 Frankfurt am Main, Germany}
\author{Yu. B. Ivanov}\thanks{e-mail: Y.Ivanov@gsi.de}
\affiliation{National Research Centre "Kurchatov Institute" (NRC "Kurchatov Institute"), 123182 Moscow, Russia}
\affiliation{National Research Nuclear University "MEPhI" (Moscow Engineering
Physics Institute), 115409 Moscow, Russia} 
\author{Iu. Karpenko}\thanks{e-mail:  karpenko@fias.uni-frankfurt.de}
\affiliation{Bogolyubov Institute for Theoretical Physics, 03680 Kiev, Ukraine}
\affiliation{INFN - Sezione di Firenze, I-50019 Sesto Fiorentino (Firenze), Italy}
%\affiliation{Frankfurt Institute for Advanced Studies (FIAS),
%Room 2|401
%Science Campus Riedberg, Ruth-Moufang-Strasse 1, 60438 Frankfurt am Main, Germany} 
%
%\author{L. Malinina}\thanks{e-mail:  ludmila.malinine@cern.ch}
%\affiliation{Veksler and Baldin Laboratory of High Energy Physics, JINR Dubna, 141980 Dubna, Russia}
%
\author{S. Merts}\thanks{e-mail:  sergey.merts@gmail.com}
\affiliation{Veksler and Baldin Laboratory of High Energy Physics, JINR Dubna, 141980 Dubna, Russia}
\author{M. Nahrgang}\thanks{e-mail:  marlene.nahrgang@phy.duke.edu}
\affiliation{Department of Physics, Duke University, Durham, North Carolina 27708-0305, USA}
\affiliation{SUBATECH, UMR 6457, Universit\'{e} de Nantes, Ecole des Mines de Nantes, IN2P3/CNRS, 4 rue Alfred Kastler, 44307 Nantes cedex 3, France}
\author{H. Petersen}\thanks{e-mail:  petersen@fias.uni-frankfurt.de}
\affiliation{Frankfurt Institute for Advanced Studies (FIAS),
Science Campus Riedberg, Ruth-Moufang-Strasse 1, 60438 Frankfurt am Main, Germany} 
\affiliation{Institut f\"ur Theoretische Physik, Goethe Universit\"at, Max-von-Laue-Strasse 1, 60438 Frankfurt am Main, Germany}
\affiliation{GSI Helmholtzzentrum f\"ur Schwerionenforschung GmbH, Planckstrasse 1, 64291 Darmstadt, Germany}
\author{O. Rogachevsky}\thanks{e-mail:  rogachevsky@jinr.ru}
\affiliation{Veksler and Baldin Laboratory of High Energy Physics, JINR Dubna, 141980 Dubna, Russia}

\begin{abstract}
We present a new event generator based on the three-fluid hydrodynamics approach for the early stage of the collision, followed by a particlization at the hydrodynamic decoupling surface
to join to a microscopic transport model, UrQMD, to account for hadronic final state interactions.
We present first results for nuclear collisions of the FAIR/NICA energy scan program 
(Au+Au collisions, $\sqrt{s_{NN}}=4-11$ GeV).
We address the directed flow of protons and pions as well as the proton rapidity distribution for two model EoS,  one with a first order phase transition the other with a crossover type softening at high densities. 
The new simulation program has the unique feature that it can describe a hadron-to-quark matter transition 
which proceeds in the baryon stopping regime that is not accessible to previous simulation programs designed for higher energies.
\end{abstract}
\date{\today}
\maketitle

%______________________________________________________________________ 
\section{Introduction}

The onset of deconfinement in relativistic heavy-ion collisions and the search for a critical endpoint 
is now in the focus of  theoretical and experimental studies of the equation of state (EoS) and the phase diagram of strongly interacting matter. 
This challenge is one of the main motivations for 
the currently running beam-energy scan at the Relativistic Heavy-Ion Collider (RHIC)
at Brookhaven National Laboratory (BNL) \cite{Stephans:2006tg} 
and at the CERN Super-Proton-Synchrotron (SPS) \cite{SPS-scan} as well as for constructing the
Facility for Antiproton and Ion Research (FAIR) in Darmstadt \cite{FAIR} and the
Nuclotron-based Ion Collider fAcility (NICA) in Dubna \cite{NICA}. 

Three-fluid hydrodynamics (3FH) \cite{Ivanov:2005yw} was derived to simulate heavy-ion collisions at moderately relativistic energies, i.e. precisely in the energy range of the 
expected onset of deconfinement. 
In recent years applications of 3FH demonstrated a strong preference of deconfinement scenarios for the explanation of available experimental data 
\cite{Ivanov:2013wha,Ivanov:2012bh,Ivanov:2013yqa,Ivanov:2013yla,Konchakovski:2014gda,Ivanov:2014zqa,Ivanov:2014ioa,Ivanov:2016sqy}. 
However, up to now 3FH has been facing certain problems. 
From the theoretical side, the concrete models lacked an afterburner stage that can 
play an important role for some observables. 
From the practical point of view, the models were not well suited for data simulations in terms of experimental events, because the model output consisted of fluid characteristics rather than of 
a set of observable particles.  

In this paper, we present first results obtained with the new Three-fluid Hydrodynamics-based Event Simulator Extended by UrQMD final State interactions (THESEUS)
and apply it to the description of heavy-ion collisions in the NICA/FAIR energy range.
This simulator provides a solution to both the above-mentioned problems.
It presents the 3FH output in terms of a set of observed particles 
and the afterburner can be run starting from this output by means of the UrQMD model 
\cite{Bass:1998ca}. 
Thus THESEUS as a new tool allows to discuss the multifaceted physics challenges at FAIR and NICA energies.
The new simulation program has the unique feature that it can describe a hadron-to-quark matter transition of first order which proceeds in the baryon stopping regime that is not accessible to previous simulation programs designed for higher energies, like QGSM or PHSD.
Besides this, with THESEUS one can address
practical questions like the influence of hadronic final state interactions and of the detector acceptance, which are necessary to understand better such that the focus can lie on important physics questions. These deal with a potential discovery and investigation of the first-order phase transition line, where during a heavy-ion collision the EoS reaches its softest point \cite{Hung:1997du}. It remains an open question how this characteristic feature of the EoS manifests itself
in observables such as flow, proton rapidity distributions and femtoscopic radii. 
Particular emphasis is on the robustness of the "wiggle" \cite{Ivanov:2015vna} in the energy scan of the midrapidity curvature of the proton rapidity distribution that has been suggested as a possible signal for a first order phase transition, expected just in the range of energies at NICA and FAIR experiments. 

At present THESEUS is not an integrated approach. 
The simulation proceeds in two steps: first, a numerical solution of the 3-fluid hydrodynamics is computed with the corresponding code. 
Based on the output of the hydrodynamic part, a Monte Carlo procedure is used to sample the ensemble of hadron distributions and the UrQMD code is engaged to calculate final state hadronic rescatterings, as will be explained below.
Another present limitation which we leave for future work is the absence of event-by-event hydrodynamic evolution. 
Therefore later by an event we mean a Monte Carlo sampled set of final hadrons, which correspond to the same (average) hydrodynamic evolution.

Beyond the scope of this paper but of interest for future research are nonequilibrium effects at the first-order phase transition due to nucleation and spinodal decomposition \cite{Csernai:1995zn,Zabrodin:1998dk,Keranen:2002sw,Nahrgang:2011vn}. 
It is expected to observe large density inhomogeneities, droplet formation and an amplification of low-momentum modes \cite{Mishustin:1998eq,Randrup:2009gp,Randrup:2010ax,Steinheimer:2012gc, Herold:2013bi,Steinheimer:2013gla,Herold:2013qda,Steinheimer:2013xxa,Nahrgang:2016eou} as a consequence. 
In order to be able to deal with these higher-order effects it is of greatest importance to correctly treat effects on the level of the EoS and sources of fluctuations like the initial and the final state, as well as experimental constraints like finite acceptance.

This paper is organized as follows. In sect. \ref{THESEUS}  a brief survey of the components of the event generator is presented: the 3FH model in subsect.~\ref{3FD}, the particlization procedure in 
subsect.~\ref{Particlization} and the UrQMD model used for afterburner simulations in subsect.~\ref{UrQMD}.
In sect. \ref{Applications} some applications of the event generator are 
presented whereby detailed plots for the energy scans of directed flow for protons and pions, 
of proton rapidity distributions and of the influence of detector acceptance and collision centrality on the baryon stopping signal are given in the appendices A, B and C, respectively. 
The conclusions are drawn in sect.~\ref{Conclusions}. 

\section{Description of the event generator THESEUS}
\label{THESEUS}
%______________________________________________________________________ 
\subsection{The 3FH model} 
\label{3FD}

The 3FH model treats the collision process from the 
very beginning, i.e. from the stage of cold nuclei, up to the particlization   
from the fluid dynamics. 
This model is a straightforward extension of the two-fluid model with radiation
of direct pions \cite{Mishustin:1988mj,Mishustin:1991sp}
and of the (2 + 1)-fluid model of 
Refs.~\cite{Katscher:1993xs,Brachmann:1997bq}.
The 3-fluid approximation is a minimal way to simulate the finite
stopping power at the initial stage of the collision. 
Within the 3-fluid approximation a generally nonequilibrium distribution of baryon-rich
matter is simulated by counter-streaming baryon-rich fluids
initially associated with constituent nucleons of the projectile (p)
and target (t) nuclei. 
Therefore, the initial conditions for the fluid evolution are two Lorentz contracted spheres with radii of corresponding nuclei and zero diffuseness, baryon density $n_B = 0.15$ fm$^{-3}$ and energy density $m_N n_B \simeq 0.14$ GeV/fm$^3$.
In addition, newly produced particles, populating
the mid-rapidity region, are associated with a fireball (f) fluid. 
Each of these fluids is governed by conventional hydrodynamic
equations.  
The continuity equations for the baryon charge read 
   \begin{eqnarray}
   \label{eq8}
   \partial_{\mu} J_{\alpha}^{\mu} (x) &=& 0,
   \end{eqnarray}
for $\alpha=$p and t, where
$J_{\alpha}^{\mu}=n_{\alpha}u_{\alpha}^{\mu}$ is the baryon
current defined in terms of proper (i.e. in the local rest frame) net-baryon density $n_{\alpha}$ and
 hydrodynamic 4-velocity $u_{\alpha}^{\mu}$ normalized as
$u_{\alpha\mu}u_{\alpha}^{\mu}=1$. Eq.~(\ref{eq8}) implies that
there is no baryon-charge exchange between p-, t- and f-fluids, as
well as that the baryon current of the fireball fluid is
identically zero, $J_{\scr f}^{\mu}=0$,  by construction. 
Equations of the energy--momentum exchange between fluids are formulated
in terms of energy--momentum tensors $T^{\mu\nu}_\alpha$ of the fluids
   \begin{eqnarray}
   \partial_{\mu} T^{\mu\nu}_{\scr p} (x) &=&
-F_{\scr p}^\nu (x) + F_{\scr{fp}}^\nu (x),
   \label{eq8p}
\\
   \partial_{\mu} T^{\mu\nu}_{\scr t} (x) &=&
-F_{\scr t}^\nu (x) + F_{\scr{ft}}^\nu (x),
   \label{eq8t}
\\
   \partial_{\mu} T^{\mu\nu}_{\scr f} (x) &=&
- F_{\scr{fp}}^\nu (x) - F_{\scr{ft}}^\nu (x)
\cr
&+&
\int d^4 x' \delta^4 \left(\vphantom{I^I_I} x - x' - U_F
(x')\tau_f\right)
\cr
&\times&
 \left[F_{\scr p}^\nu (x') + F_{\scr t}^\nu (x')\right],
   \label{eq8f}
   \end{eqnarray}
where the $F^\nu_\alpha$ are friction forces originating from
inter-fluid interactions. $F_{\scr p}^\nu$ and $F_{\scr t}^\nu$ in
Eqs.~(\ref{eq8p})--(\ref{eq8t}) describe energy--momentum loss of the 
baryon-rich fluids due to their mutual friction. A part of this
loss $|F_{\scr p}^\nu - F_{\scr t}^\nu|$ is transformed into
thermal excitation of these fluids, while another part $(F_{\scr
p}^\nu + F_{\scr t}^\nu)$ gives rise to particle production into
the fireball fluid (see Eq.~(\ref{eq8f})). $F_{\scr{fp}}^\nu$ and
$F_{\scr{ft}}^\nu$ are associated with friction of the fireball
fluid with the p- and t-fluids, respectively. 
Here $\tau_f$ is the formation time, and
   \begin{eqnarray}
   \label{eq14}
U^\nu_F (x')=
\frac{u_{\scr p}^{\nu}(x')+u_{\scr t}^{\nu}(x')}%
{|u_{\scr p}(x')+u_{\scr t}(x')|}
   \end{eqnarray}
is the  4-velocity of the free propagation of the produced fireball 
matter.
In fact, this is the velocity of the fireball matter at the moment of its production. 
According to Eq.~(\ref{eq8f}),  this matter gets formed only 
after the time span $U_F^0\tau_f$ upon the production, and in a 
different space point ${\bf x}' - {\bf U}_F (x') \ \tau_f$, as
compared to the production point ${\bf x}'$. 
 The friction between fluids
was fitted to reproduce the stopping power observed in proton
rapidity distributions for each EoS, as it is described in Refs. 
\cite{Ivanov:2005yw,Ivanov:2013wha} in detail.

\begin{figure}[!hbt]
%\vspace*{-14mm}
\includegraphics[width=0.45\textwidth]{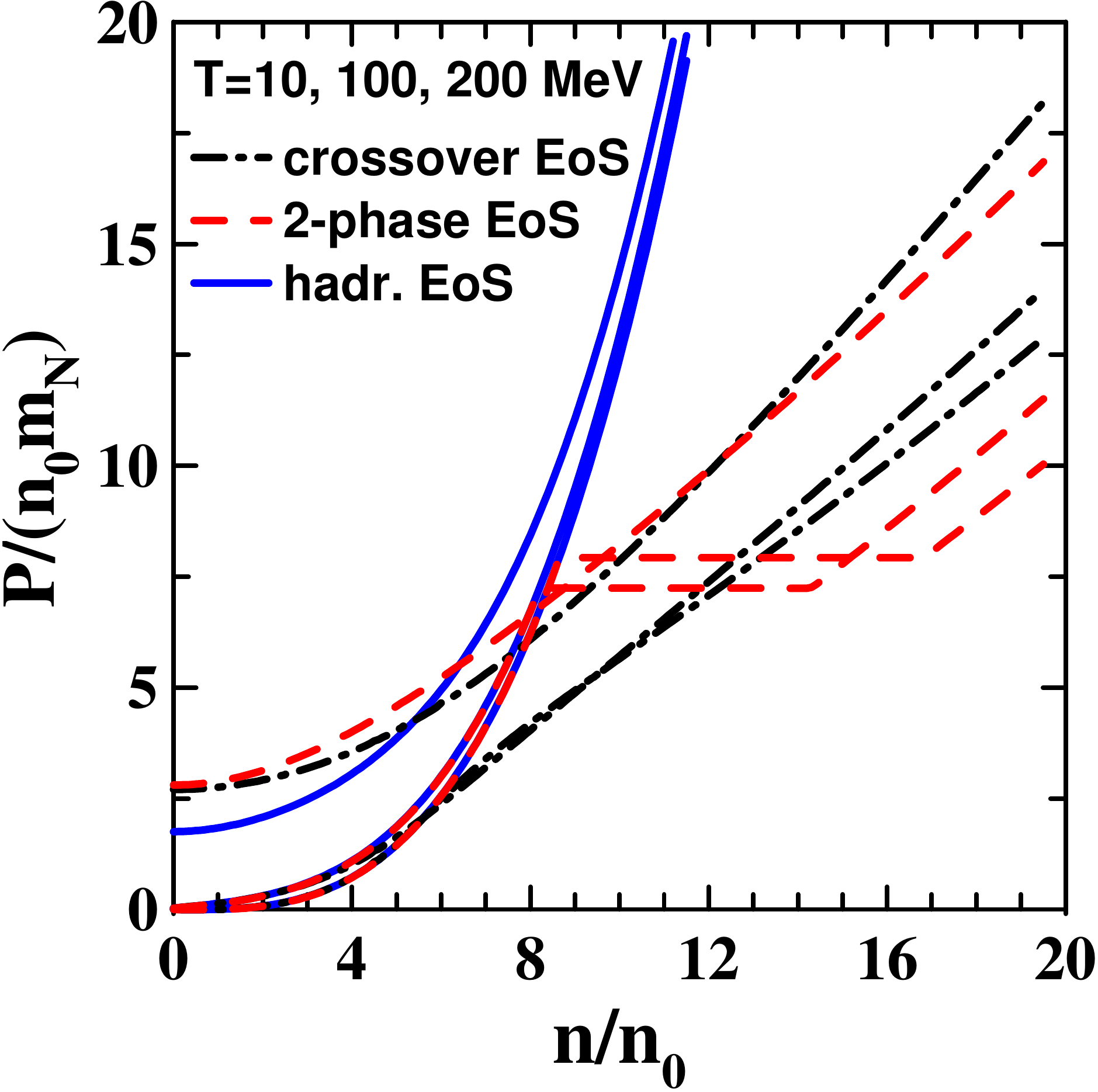}
%\vspace*{-20mm}
 \caption{{
Pressure scaled by the product of normal nuclear density ($n_0=$ 0.15 fm$^{-3}$) and 
nucleon mass ($m_N$) versus baryon density scaled by the normal nuclear density
for three considered equations of state. Results are presented for three different
temperatures $T=$ 10, 100 and 200 MeV (from bottom upwards for corresponding curves).  
}} 
\label{fig1}
\end{figure}

Different equations of state (EoS) can be applied within the 3FH model. 
The recent series of simulations 
\cite{Ivanov:2013wha,Ivanov:2012bh,Ivanov:2013yqa,Ivanov:2013yla,Konchakovski:2014gda,Ivanov:2014zqa,Ivanov:2014ioa,Ivanov:2016sqy}
was performed  
employing three different types of EoS: a purely hadronic EoS   
\cite{gasEOS} (hadr. EoS) and two versions of the EoS involving    
deconfinement  \cite{Toneev06}. 
The latter two versions are an EoS with a first-order phase transition (2-phase EoS) 
and one with a smooth crossover transition (crossover EoS). 
%{\color{red}
The hadronic EoS is well in agreement with knowns constraints \cite{Klahn:2006ir},
in particular the flow constraint by Danielewicz et al. \cite{Danielewicz:2002pu},
see \cite{Bastian:2015avq} for an explicit comparison.  
%}
Figure \ref{fig1} illustrates the differences between the three considered EoS. 
\\[5mm]

The numerical scheme of
the code is based on the modified particle-in-cell method \cite{Roshal81} 
which is an extension of the scheme first applied in Los Alamos \cite{Harlow76}.
In the particle-in-cell method the matter is represented by
an ensemble of Lagrangian test particles. 
They are used for calculation of the drift transfer of the baryonic charge, energy,
and momentum. 
In the present scheme the test particle has the size of the cell. 
Therefore, when a single test particle is moved on the grid, it changes quantities 
in eight cells, with which it overlaps. 
These spatially extended particles make the scheme smoother and hence more stable. 
The transfer because of pressure gradients, friction between fluids and production of
the fireball fluid, is computed on the fixed grid (so called Euler step of the scheme). 
Simulations are performed in the frame of equal velocities of colliding nuclei. 
The numerical-scheme input of the present 3FH calculations is
described in detail in Ref.~\cite{Ivanov:2005yw}.
   
\begin{figure*}[!ht]
%\vspace*{-14mm}
\includegraphics[width=18.0cm]{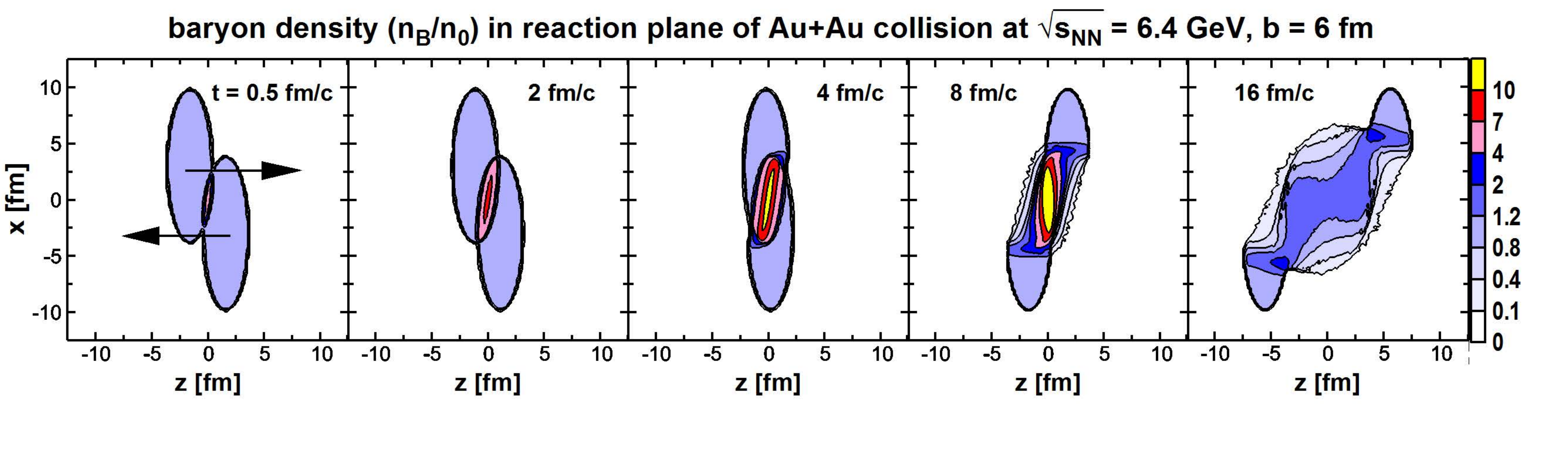}
%\vspace*{-20mm}
 \caption{
Evolution of the proper  
baryon density ($n_B/n_0$) scaled by the the normal nuclear density ($n_0=0.15~$fm$^{-3}$)
in the reaction plane for a semi-central ($b=6~$fm) Au+Au collision at $\sqrt{s_{NN}}=6.4~$GeV.
} 
\label{fig2}
\end{figure*}

An application of the 3FH model is illustrated in Fig.~\ref{fig2} 
where the evolution of the proper (i.e. in the local rest frame
obtained by diagonalization of the energy-momentum tensor, 
see the next subsection) 
baryon density in the reaction plane is presented for a semi-central 
(impact parameter $b=$ 6 fm) Au+Au collision at $\sqrt{s_{NN}}=$ 6.4 GeV
($E_{\rm lab}= 20 $ A GeV).  
The simulation was performed with the crossover 
EoS without freeze-out. 
As can be seen from that figure, very high baryon densities are reached 
in the central region of the colliding system.

The freeze-out criterion used in the 3FH model is $\varepsilon < \varepsilon_{\scr{frz}}$, 
where $\varepsilon$ is the total energy density of all three fluids in their common rest 
frame.
More details can be found in Refs.~\cite{71,74}
The freeze-out energy density $\varepsilon_{\scr{frz}}=0.4~$GeV/fm$^3$ 
was chosen mostly on the condition of the best reproduction 
of secondary particle yields for all considered scenarios, see \cite{Ivanov:2005yw}.  
An important feature of the 3FH freeze-out is an antibubble prescription, preventing the formation of
bubbles of frozen-out matter inside the dense matter while it is still hydrodynamically evolving. 
The matter is allowed to be frozen out only if 
it is located near the border with the vacuum (this piece of matter gets locally frozen out). 
The thermodynamic quantities of the frozen-out
matter are recalculated from the in-matter EoS, with which
the hydrodynamic calculation runs, to the hadronic gas EoS%
\footnote{
In this gas EoS 48 different hadronic
species are taken into account. Each hadronic species includes all the
relevant isospin states; e.g., the nucleon species includes protons and
neutrons.}.
This is done because a part of the energy is still accumulated in
collective mean fields at the freeze-out instant. 
This mean-field energy needs to be released before entering the hadronic cascade 
in order to facilitate energy conservation. 

The output of the model is recorded in terms of Lagrangian test particles (i.e. fluid droplets)
for each fluid $\alpha$ (= p, t or f).
Each particle contains information on space-time coordinates ($t,\bf{x}$) of the frozen-out matter, 
proper volume of the test particle of the $\alpha$ fluid ($V_\alpha^{\rm pr}$), 
hydrodynamic velocity ($u^{\mu}_\alpha$) in the frame of computation, 
temperature ($T_{\alpha}$), baryonic ($\mu_{\rm B\alpha}$)  and strange ($\mu_{\rm S\alpha}$) chemical potentials.

\subsection{Particlization}
\label{Particlization}

In the multi-fluid approach one simulates the heavy ion collision from its very first moment using fluid dynamics. 
However, once the system becomes too dilute, the fluid approximation loses its applicability and individual particles are the relevant degrees of freedom. 
The process of changing from a fluid to a particle description is called "particlization" \cite{Huovinen:2012is}. 
Since we supplement the 3FH with a hadronic cascade, the particlization is not freeze-out anymore.
By definition there are only resonance decays after freeze-out, whereas in the present generator final state hadronic rescattering processes are simulated as well using the UrQMD code.

The particlization criterion is chosen to be the same as freeze-out criterion in \cite{Ivanov:2005yw}, e.g.,
$$\varepsilon_{\rm tot}<\varepsilon_{\rm frz},$$
where $\varepsilon_{\rm tot}$ is defined as:
$$\varepsilon_{\rm tot}=T^{*00}_{p}+T^{*00}_{t}+T^{*00}_{f}$$
and the asterisk denotes a reference frame where the nondiagonal components of the total energy momentum tensor are zero. 
This choice allows to study the influence of hadronic rescatterings to the observables by comparing them with the ones calculated in previous 3-fluid hydrodynamic models.

For the details of fluid to particle conversion the reader is referred to \cite{Ivanov:2005yw}, whereas here we repeat the details important for the construction of the Monte Carlo sampling procedure. 
Both the baryon-rich projectile and target fluids as well as the fireball fluid are being frozen out in small portions, therefore the output of the particlization procedure is a set of droplets (or surface elements). Each droplet is characterized by its proper volume $V^{\rm pr}$, temperature $T$, baryon, $\mu_{\rm B}$, strange chemical potentials $\mu_{\rm S}$, 
and the collective flow velocity $u^\mu$.

The thermodynamic parameters of the droplets correspond to a free hadron resonance gas. 
Therefore, we proceed with sampling the hadrons according to their phase space distributions 
(see Eq.~(33) in \cite{Ivanov:2005yw}), which are expressed in the rest frame of the fluid element (FRF) 
as
\begin{equation}\label{eq-momentum-rf}
p^{*0} \frac{d^3 N_i}{d^3 {\bf p^*}}=\sum\limits_{\alpha} \frac{g_i V_\alpha^{\rm pr}}{(2\pi)^3}
 \frac{p^{*0}}{\exp\left[ (p^{*0} - \mu_{\alpha i}) / T_\alpha \right] \pm 1}
\end{equation}
where the asterisk denotes momentum in the fluid rest frame, where $u^{*\mu}_\alpha=(1,0,0,0)$, 
$\mu_{\alpha i}=B_i\cdot\mu_{\alpha\rm B} + S_i\cdot\mu_{\alpha\rm S}$ is the chemical potential of  hadron $i$ with baryon number $B_i$, strangeness $S_i$, degeneracy factor  $g_i$, 
and the $\alpha$ summation runs over droplets from all (p, t and f) fluids.

\begin{figure*}[!th]
\includegraphics[scale=0.85]{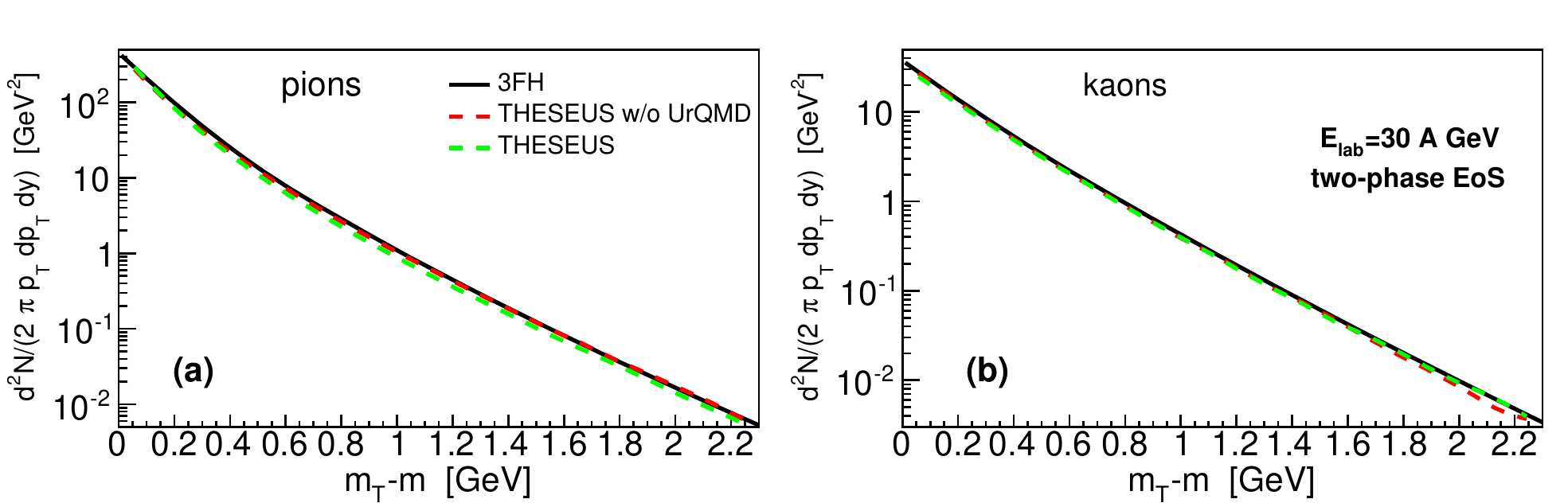}
\caption{(Color online) Transverse momentum spectrum for pions (a) and kaons (b) for  central Au+Au collisions ($b=2~$fm) at $E_{\rm lab}=30~$A~GeV for the 2-phase EoS. 
Comparison between results from the 3FH model (black solid lines) and THESEUS without UrQMD (red dashed lines) show excellent agreement.
Comparing these results with the full THESEUS result (green dashed line) shows that the UrQMD hadronic rescattering leads to a slight steepening of the pion $p_T$ spectrum.
\label{fig-pt}}
\end{figure*}

\begin{figure*}[!th]
\includegraphics[scale=0.9]{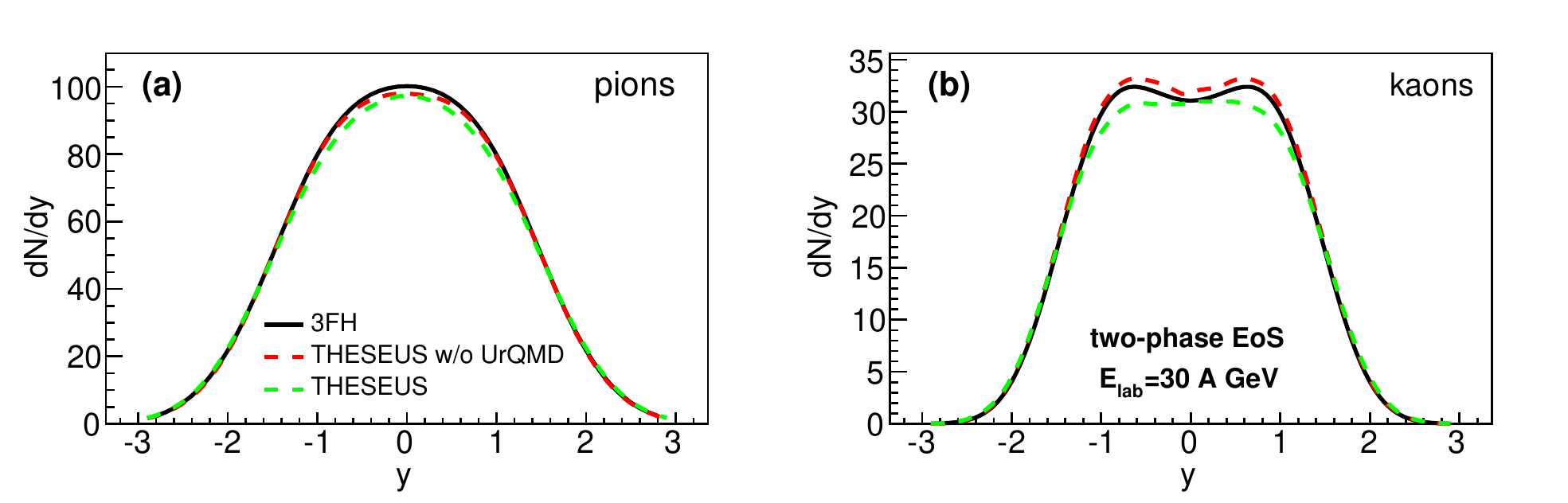}
\caption{(Color online) Rapidity distribution for pions (a) and kaons (b) for central Au+Au collisions ($b=2~$fm) at $E_{\rm lab}=30~$A~GeV for the 2-phase EoS. 
Comparison between results from the 3FH model (black solid lines) and THESEUS without UrQMD (red dashed lines) show excellent agreement.
Comparing these results with the full THESEUS result (green dashed line) shows that the UrQMD hadronic rescattering smeares out the double-peak structure in the kaon rapidity spectrum.
\label{fig-dndy}}
\end{figure*}

The use of temperature and chemical potentials implies a grand canonical ensemble for each surface element.
The sampling is therefore organized as a loop over all droplets, every iteration of which consists of the following steps \cite{Karpenko:2015xea,Karpenko:2013ama}
\begin{itemize}
 \item average multiplicities of all hadron species are calculated according to
 \begin{equation}
 \label{mult}
  \Delta N_{i,\alpha}=V^{\rm pr}_\alpha n_{i,\rm th}(T, \mu_i),
 \end{equation}
 together with their sum $\Delta N_{\rm tot, \alpha}=\sum_i \Delta N_{i,\alpha}$;
 \item total (integer) number of hadrons from each surface element is sampled according to Poisson distribution with mean $\Delta N_{\rm tot, \alpha}$. If the number is greater than zero, sort of hadron is randomly chosen based on probabilities $\Delta N_{i,\alpha}/\Delta N_{\rm tot, \alpha}$;
 \item hadron's momentum in FRF $p^*$ is sampled according to (\ref{eq-momentum-rf}), which is isotropic in momentum space;
 \item momentum vector is Lorentz boosted to the global frame of the collision.
\end{itemize}
In the present version of the generator, also from the arguments of consistency with preceding hydrodynamic evolution, we do not apply any corrections over the grand canonical procedure to account for effects of charge or energy conservation. 
Therefore, particle multiplicities fluctuate from event to event according to the composition of grand canonical ensembles given by the individual droplets.

\subsection{UrQMD simulation of final state interactions}
\label{UrQMD}

The Ultra-relativistic Quantum Molecular Dynamics (UrQMD) approach \cite{Bass:1998ca}
treats hadrons and resonances up to a mass of $\sim 2.2$ GeV. 
All binary interactions are treated via the excitation and decay of resonances or string 
excitation and decay and elastic scatterings. 
It is crucial for a state-of-the-art event generator to treat the interactions during the late non-equilibrium 
hadronic stage of heavy ion reactions properly. 
At RHIC and LHC notable differences in the proton yields have been observed and the identified particle spectra and flow observables show an effect of the hadronic rescattering 
(for a review of hybrid approaches see \cite{Petersen:2014yqa}). 
At lower beam energies as they are investigated in this work, the hadronic stage of the reaction is of utmost importance. 
In \cite{Auvinen:2013sba} it has been shown, that the excitation function of elliptic and triangular flow can only  be understood within a combined hydrodynamics+transport approach. 
UrQMD constitutes an effective solution of the relativistic Boltzmann equation and therefore provides 
access to the full phase-space distribution of all individual particles at all times. In this work the effect of hadronic rescattering in the final state
on the identified particle spectra and the rapidity dependent directed flow is demonstrated in detail. 
\section{Results}
\label{Applications}
In this section we present a selection of first results from THESEUS for the energy scan ($\sqrt{s_{NN}}=4-11$ GeV) planned at the NICA-MPD collider experiment, which has overlap with the energy range that will become accessible in the fixed target experiments at FAIR-CBM, see table \ref{tab:energies}. 

\begin{table}[!h]
\begin{tabular}{l|c|c|c|c|c|c|c}
\hline \hline
$\sqrt{s_{NN}}$ [GeV] & 4.3 & 4.7 & 5.6 & 6.4 & 7.7& 9.2 & 11.6
\\
\hline
$E_{\rm lab}$ [A~GeV] & 8 & 10 & 15 & 20 & 30 & 43 & 70 \\
\hline
\hline
\end{tabular}
\caption{Center of mass energies $\sqrt{s_{NN}}$ for the NICA-MPD energy scan with Au+Au collisions (upper row) and their equivalent fixed target energies in the laboratory system (lower row).
\label{tab:energies}
}
\end{table}
\subsection{Tests of the particlization routine: Spectra of pions, kaons and protons}
We start by showing the transverse momentum distributions of pions ( $(\pi^+ + \pi^0 + \pi^-)/3$) and kaons ($(K^+ + K^0)/2$) in Fig.~\ref{fig-pt}. 
They are calculated from a sample of 30000 events generated according to the Monte Carlo procedure described above, and are compared in the plot to 
%{\color{red}
3FH and THESEUS w/o UrQMD, where "3FH" corresponds to the model in which particle spectra are obtained from  direct integration of Eq.~(\ref{eq-momentum-rf}), and "THESEUS w/o UrQMD" means that particles are obtained from Monte-Carlo sampling of Eq.~(\ref{eq-momentum-rf}).
%} 
The 3FH evolution simulates Au+Au collisions at $E_{\rm lab}=30$~A~GeV with the two-phase EoS. 
We observe excellent agreement up to $p_T=2.2$~GeV, which is limited by the generated event statistics.
In Fig.~\ref{fig-dndy} we show the rapidity distributions for the same setup. 
The rapidity distributions reveal a small difference in kaon yields, and an even smaller one for pions, which is attributed to differences in the large mass sector of the resonance tables and branching ratios. Nevertheless, the shapes of rapidity distributions agree beautifully.
In Figs.~\ref{fig-pt} and \ref{fig-dndy} we show also the effect of the UrQMD hadronic final state interactions
which are included in THESEUS. They lead to a slight steepening of the $p_T$ spactrum for pions and to a
reduction of the double-peak structure in the kaon rapidity spectrum. 
Both are sufficiently gentle effects to not spoil our conclusion.
The tests demonstrate that both the procedure of particle sampling
at particlization and the resonance decay kinematics are implemented correctly.
\subsection{Directed flow of protons and pions}
Next we test whether more subtle features of particle distributions are preserved by the particlization procedure and how they are affected by the hadronic cascade. 
First we calculate the directed flow coefficient $v_1$ for pions and protons as a function of rapidity using the reaction plane method
$$v_1(y)=\left< \cos(\phi-\Psi_{\rm RP}) \right>=\left< p_x/\sqrt{p_x^2+p_y^2} \right>,$$
where $\Psi_{\rm RP}=0$ in the model, since the impact parameter is always directed along x-axis.
Although the generator makes it possible to apply different methods of flow analysis over generated events, we use the reaction plane method in order to perform a one-to-one comparison between results from 
THESEUS with and without UrQMD and the corresponding ones from the basic 3FH model.

\begin{figure}[!th]
\label{proton_v1_slope}
\includegraphics[width=0.45\textwidth]{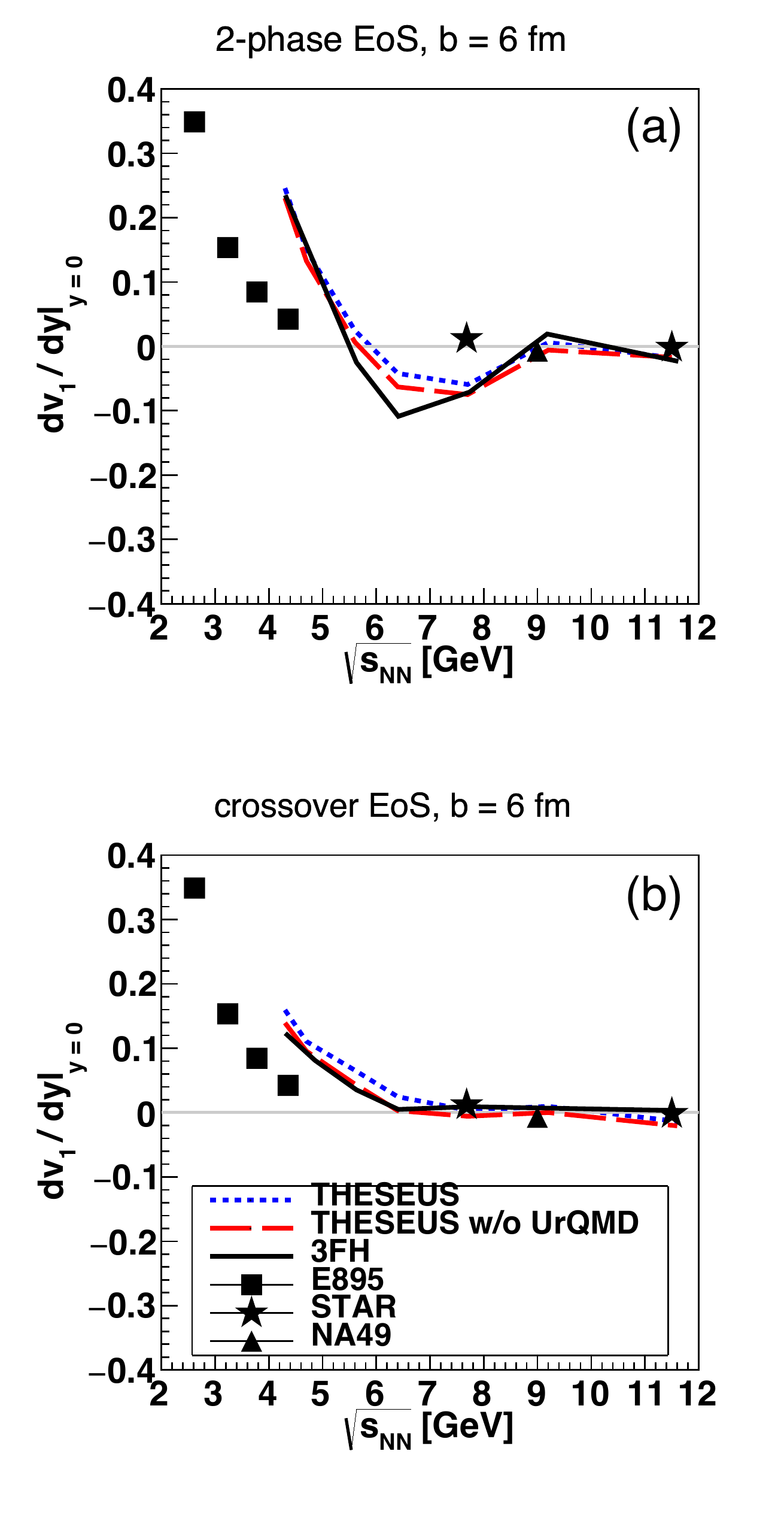}
\caption{(Color online) 
Energy scan of the slope of the directed flow ($dv_1/dy$) of protons for 
semicentral ($b=6$ fm) 
Au+Au collisions.  
We compare results for 3FH (black solid line), THESEUS (blue short-dashed line)
and THESEUS without UrQMD hadronic rescattering (red long-dashed line)
for the 2-phase EoS (a) and the crossover EoS (b). 
Data from the AGS experiment E895 
\cite{Liu:2000am}
are shown by filled squares, data from the STAR beam energy scan 
\cite{Adamczyk:2014ipa}
are given by star symbols and a data point data from NA49 
\cite{Alt:2003ab}
by a filled triangle.
\label{fig-proton_v1_slope}}
\end{figure}

\begin{figure}[!th]
\includegraphics[width=0.45\textwidth]{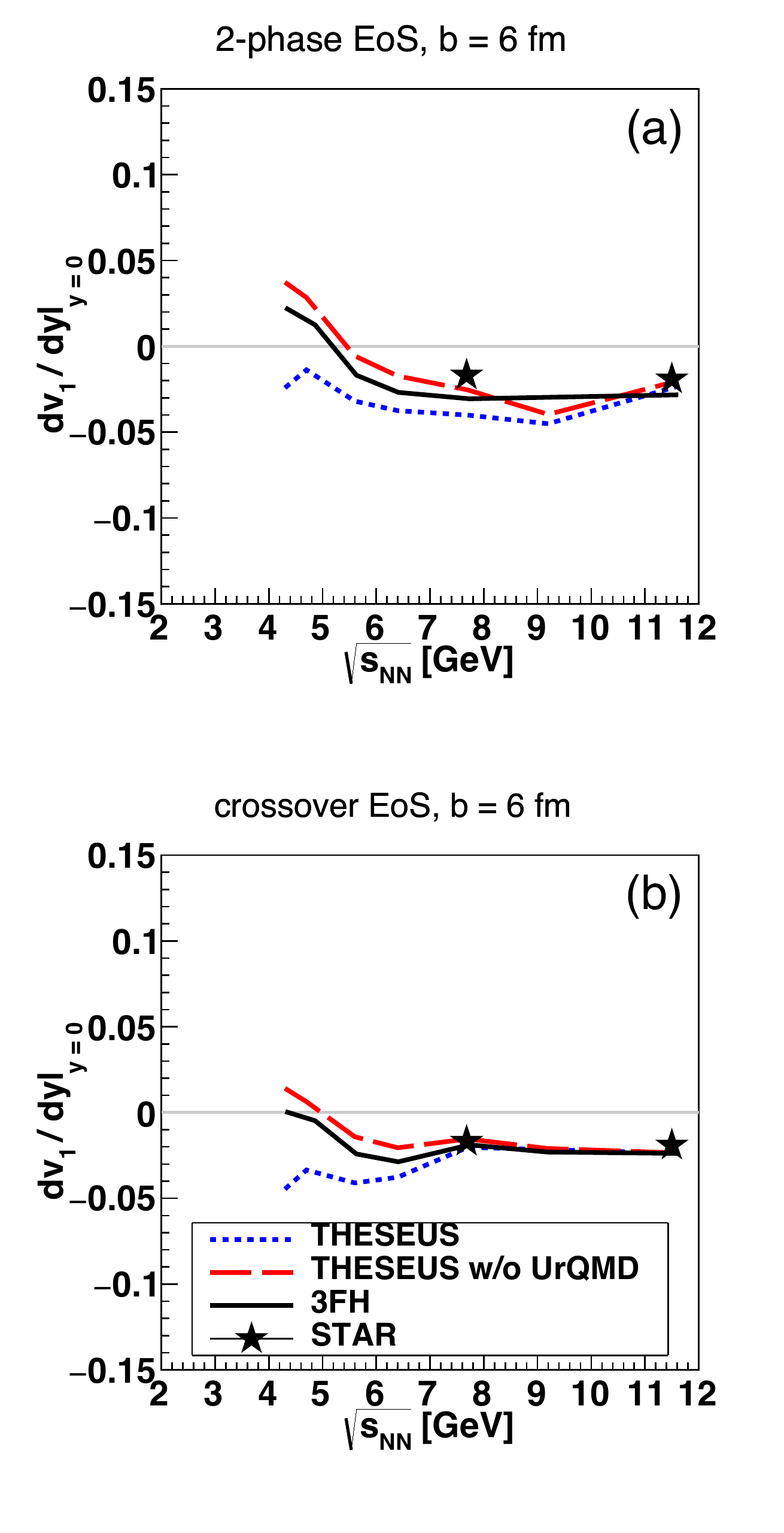}
\caption{(Color online) Energy scan of the slope of the directed flow ($dv_1/dy$) of pions for 
semicentral ($b=6$ fm) Au+Au collisions.  
We compare results for 3FH (black solid line), THESEUS (blue short-dashed line)
and THESEUS without UrQMD hadronic rescattering (red long-dashed line)
for the 2-phase EoS (a) and the crossover EoS (b). 
Data from the STAR beam energy scan \cite{Adamczyk:2014ipa}
are shown by star symbols.
\label{fig-pion_v1_slope}}
\end{figure}

The rapidity dependent directed flow $v_1(y)$ of protons and pions for different collision energies, impact parameters and EoS can be found in Appendix \ref{app:A}, while the
net proton rapidity distributions for different collision energies and EoS are given in Appendix \ref{app:B}.
In Figs.~\ref{fig-proton_v1_slope} and \ref{fig-pion_v1_slope} we present the distributions  in a condensed form using the slope of the directed flow at midrapidity $dv_1/dy$
%{\color{red}
calculated in the interval $|\Delta y|<0.5$ around midrapidity.
%}
Dashed lines show the results from THESEUS without hadronic cascade, where we quantitatively reproduce the results from basic 3FH model, including the dip in the $dv_1/dy$ of (net-)protons in semi-central events for the EoS with a first-order phase transition denoted as "2-phase EoS".
We would like to note that a similar flow pattern appears also for the light nuclear clusters such as deuterons \cite{Bastian:2016xna}.

Turning the UrQMD hadronic cascade on for the final state (dotted lines in Figs.~\ref{fig-proton_v1_slope} and \ref{fig-pion_v1_slope} ) we observe that the cascade has only a small effect on the excitation function of the proton $dv_1/dy$.

However, for the pions the hadronic cascade changes flow to antiflow at low energies.
This behaviour can be understood as follows.  
If there is only hydrodynamics, the pions are emitted along the fluid flow, while when there is rescattering they are blocked by the baryonic matter in the projectile and target region, therefore the anti-flow appears. 
This was first demonstrated in Ref.~\cite{Bass:1993em}. 
This effect of the pion shadowing is more spectacular in Fig.~\ref{8AGeV} where the directed flow 
of protons and pions at $E_{\rm lab}=8~$A~GeV is presented. As seen, the proton $v_1$ is 
practically insensitive to the UrQMD afterburner, while the pion $v_1$ is strongly affected 
by this afterburner. The afterburner even changes the pion $v_1$ flow to an  antiflow. 
The effect of the pion shadowing becomes weaker with the collision energy rise, 
as it is seen from Fig.~\ref{30AGeV}, because the midrapidity region becomes 
less baryon abundant. 
Though, at larger collision energies and peripheral rapidities, 
this shadowing is still noticeable. 
%This shadowing results in better agreement with the STAR data on pion $v_1$ 
As can be seen from Fig.~\ref{fig-pion_v1_slope}, the most dramatic effect of the UrQMD hadronic rescattering is the prediction that pion antiflow persists for energies below the present limit of the  STAR beam energy scan data \cite{Adamczyk:2014ipa} at $\sqrt{s_{NN}}=7.7$ GeV, for both EoS cases: first-order and crossover phase transition.

\begin{figure}[!h]
\includegraphics[width=0.45\textwidth]{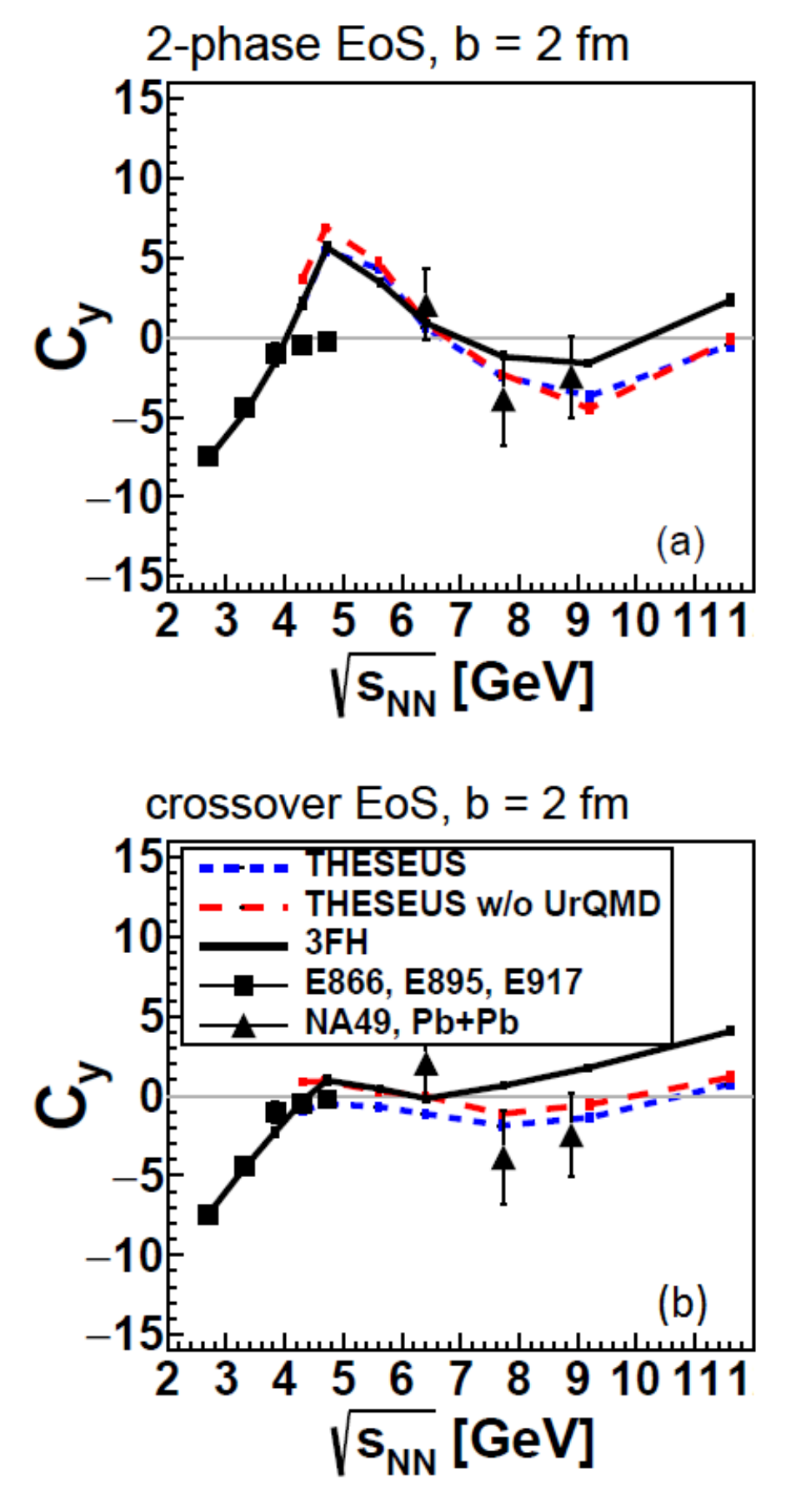}
\caption{(Color online) 
Energy scan for the curvature $C_y$ of the net proton rapidity distribution at midrapidity for 
central Au+Au collisions with impact parameter $b=2~$fm. 
We compare the 3FH model result (black solid lines) with THESEUS (blue short-dashed lines) and 
THESEUS without UrQMD (red long-dashed lines).
The results for the two-phase EoS (a) are compared to those for the crossover EoS (b).  
The "wiggle" as a characteristic feature for the EoS with a first order phase transition is rather robust against  hadronic final state interactions.
Data from AGS experiments are shown by filled squares, data from NA49 by filled triangles.
\label{stopping}
}
\end{figure}

\subsection{Baryon stopping signal for a first-order phase transition}
In Fig.~\ref{stopping} we show the reduced curvature of the net proton rapidity distribution 
(see App.~\ref{app:B} for the simulation of the energy scan of the net proton rapidity distribution itself)
$C_y=y_{\rm cm}^2 (d^3 N_{\rm net-p}/dy^3) / (dN_{\rm net-p}/dy)$, where $y_{\rm cm}$ 
is the rapidity of the center of mass of the colliding system in the frame of the target 
\cite{Ivanov:2010cu,Ivanov:2011cb,Ivanov:2015vna}.
Because of a narrower collision energy range chosen here, we observe only the peak-dip part of the
so-called ``peak-dip-peak-dip'' structure reported in \cite{Ivanov:2010cu,Ivanov:2011cb,Ivanov:2015vna}.
The reduced curvature is calculated by fitting the rapidity distribution with a 2$^{\rm nd}$ order polynomial of the form $P_2(y)=ay^2+by+c$ for which then $C_y=y^2_{\rm beam} 2a/c$ results.

\begin{figure}[!h]
\includegraphics[width=0.45\textwidth]{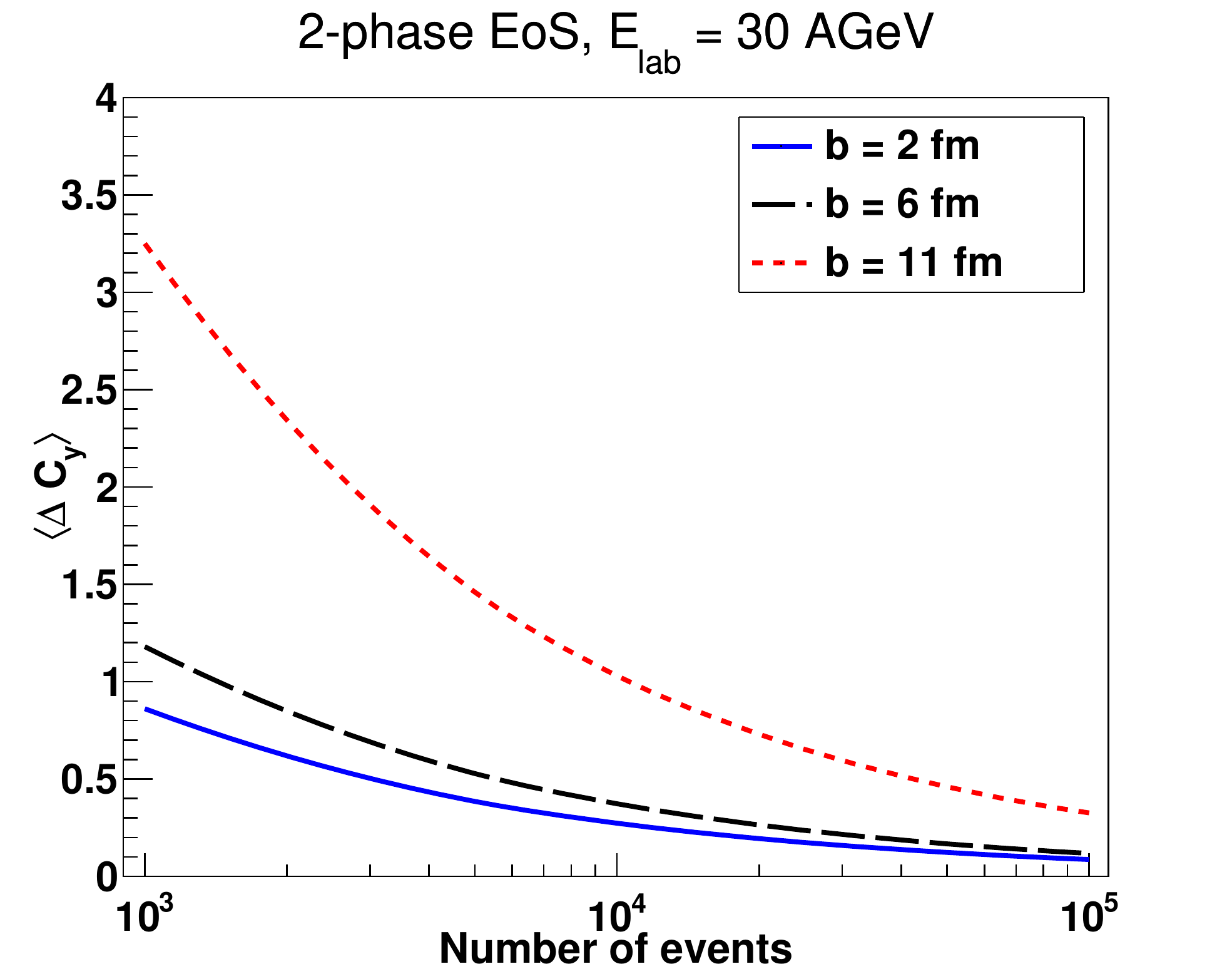}
\caption{(Color online) 
Mean standard deviation ("error") for the curvature as a function of the event statistics for central 
($b=2~$fm), semicentral ($b=6~$fm) and peripheral ($b=11~$fm) Au+Au collisions at 
$E_{\rm lab}=30~$A~GeV for the 2-phase EoS model.}
\label{fig-DeltaCy}
\end{figure}

Contrary to the basic 3FH model which can calculate $C_y$ with any given precision, in the Monte Carlo procedure the accuracy depends on the event statistics and binning. 
The error for $C_y$ can be expressed as
$$
\Delta C_y = \frac{2y^2_{\rm beam}}{c}\sqrt{(\Delta a)^2 + \frac{a^2}{c^2}(\Delta c)^2},
$$
where $\Delta a$ and $\Delta c$ denote the statistical errors of the parameters $a$ and $b$ 
from the fit to the generator output. 
The dependence of $\Delta C_y$ on the number of events is shown in Fig.~\ref{fig-DeltaCy}. 
Given that $C_y$ itself has a values no larger than several units in the collision energy range under consideration, one can conclude that a reliable determination of $C_y$ requires not less than $10^4$ events for central and semi-central collisions and $10^5$ events for peripheral collisions. Larger required statistics for peripheral events is a consequence of the lower average event multiplicity.

The robustness of the baryon stopping signal for a first-order phase transition against experimental cuts in the $p_T$ acceptance has been discussed in \cite{Ivanov:2015vna}. In App.~\ref{app:C} we provide results for these cuts with and without the UrQMD hadronic cascade, for three centralities. 
We demonstrate that the baryon stopping signal is robust against hadronic rescattering.  

\begin{figure}[!th]
\includegraphics[width=0.45\textwidth]{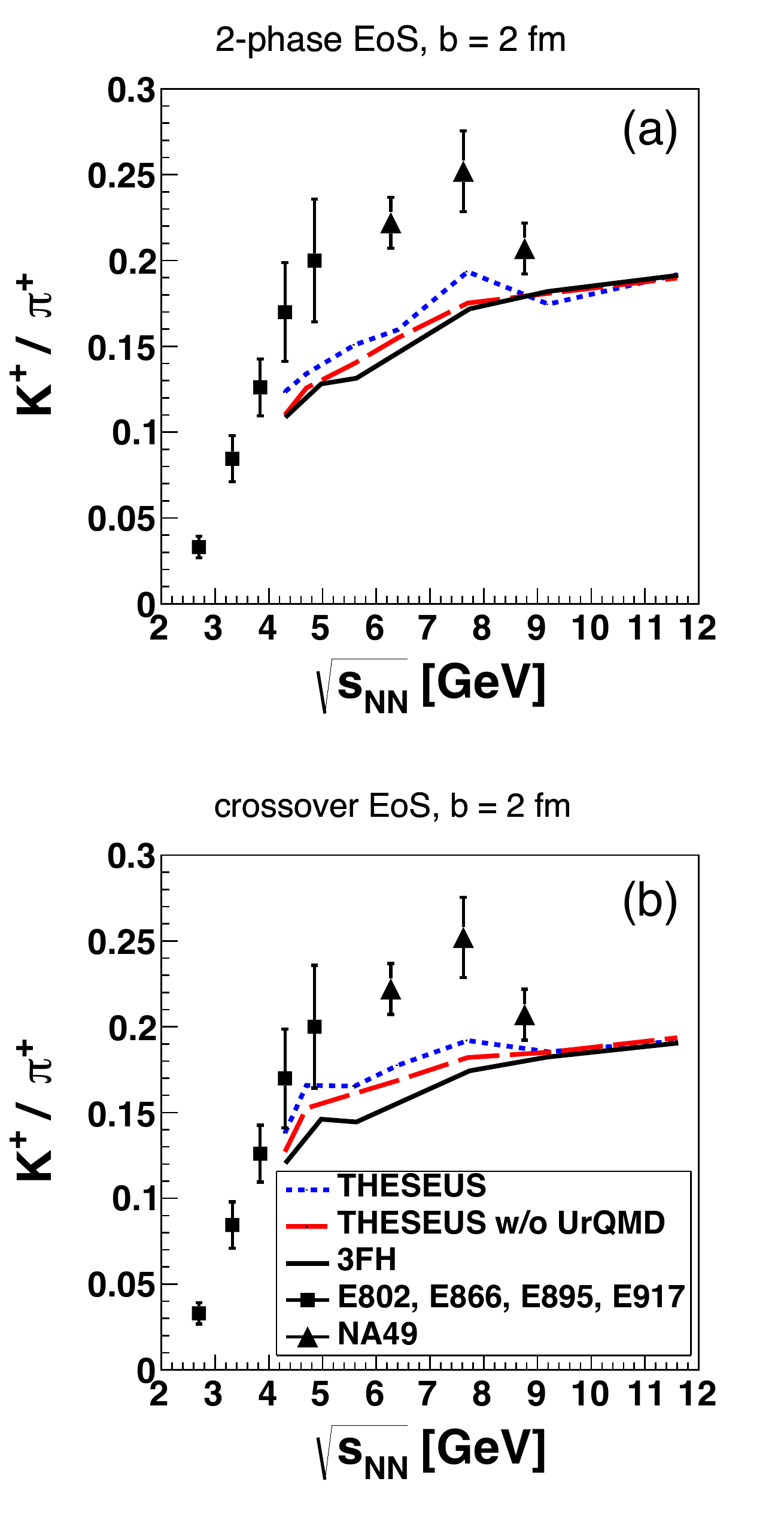}
\caption{(Color online)
Energy scan for the particle ratio $K^+/\pi^+$ in the NICA energy range for 
central Au+Au collisions (impact parameter $b=2~$fm)
with (blue lines) and without (red lines) the UrQMD hadronic rescattering.
The calculation with a first order phase transition in the EoS (a) is compared to that with the crossover EoS (b). 
For comparison we show the results without particlization and UrQMD rescattering and experimental data, taken from Fig.~11 of Ref.~\cite{Ivanov:2013yqa}.
Data from AGS experiments are shown by filled squares, data from NA49 by filled triangles.
\label{fig-horn}}
\end{figure}

\subsection{The "horn" effect?}
In Fig.~\ref{fig-horn} we show the $K^+/\pi^+$ ratio. 
The comparison between basic 3FH calculations (black solid lines) and THESEUS with UrQMD switched off shows satisfactory correspondence.
Some differences between the curves especially at lower energies can be traced back to smaller differences in kaon and pion yields we observed in Fig.~\ref{fig-dndy}. 
However, we find that turning the hadronic cascade on does not influence the kaon to pion ratio.
As the 3FH model itself, also THESEUS in its present version is not yet capable of describing the "horn" effect discovered in the NA49 data for the $K^+/\pi^+$ ratio.
It is interesting to note that aspects of the "horn" effect could be attributed to the hydrodynamical stage and the core-corona separation within the UrQMD hybrid model \cite{Steinheimer:2011mp}.
Most recently, as further aspects of the "horn" effect the chiral symmetry restoration \cite{Palmese:2016rtq}
and an anomalous $K^+$ bound state mode \cite{Dubinin:2016wvt} 
in dense matter have been pointed out.

\section{Conclusions}
\label{Conclusions}
We have assembled the new event generator THESEUS that is based on a three-fluid hydrodynamics description of the early and dense stage of the collision, followed by a particlization 
as input to the UrQMD "afterburner" accounting for hadronic final state interactions. 
 
We presented first results from THESEUS for the FAIR/NICA energy scan addressing the directed flow of protons and pions as well as the proton rapidity distribution for two model EoS, 
one with a first order phase transition the other with a crossover type softening at high densities. 
The new simulation program has the unique feature that it can describe a hadron-to-quark matter transition which proceeds in the baryon stopping regime that is not accessible to previous simulation programs that are designed for higher energies.

We have found that the hadronic cascade which is switched on after the particlization has little effect 
on the proton flow observables. 
In particular, the hadronic final state interactions preserve the characteristic non-monotonic behavior 
of the rapidity slope of directed flow of protons and the characteristic collision energy dependence
of the wiggle in the curvature of the rapidity distribution of net protons. 
However, for pions in non-central collisions at lower energies the hadronic cascade leads to a qualitative change of the emission pattern (from flow to antiflow).
The present analysis in the improved 3FH model THESEUS has demonstrated that the predicted 
baryon stopping signal for a first-order phase transition in heavy-ion collisions at NICA/FAIR energies
is a robust feature. 

The next steps planned in the development of THESEUS include the task to make it an integrated approach and to explore possible mechanisms that could explain the observed "horn" effect for the 
$K^+/\pi^+$ ratio. 
Another very interesting direction concerns the production and flow of light nuclear clusters for which 
first results look very promising \cite{Bastian:2016xna}.
 
% ____________________________________________________________________
\begin{acknowledgments} 
We thank H.~Gutbrod, A.~S.~Sorin and I.~Tserruya for the initiation of this work 
and for their continued interest in its development. 
P.~Huovinen is acknowledged for a very careful reading of the manuscript and critical remarks which 
helped improving the clarity of the text.
We are grateful to A.~S.~Khvorostukhin, V.~V.~Skokov,  and V.~D.~Toneev for providing 
us with the tabulated 2-phase and crossover EoS.
The calculations were performed at the computer cluster of GSI (Darmstadt). 
Y.B.I. and I.K. thank for support from the NICA project and for the hospitality extended to them at JINR Dubna during their visits in 2014 and 2015 when this work was initiated.
The work of D.B.  was supported in part by the Polish NCN under grant 
UMO-2011/02/A/ST2/00306, by the Hessian LOEWE initiative through HIC for FAIR
and by the MEPhI Academic Excellence Project under contract No. 02.a03.21.0005.
H.P. acknowledges funding of a Helmholtz Young Investigator Group VH-NG-822 from
the Helmholtz Association and GSI.
M.N. acknowledges support from a fellowship within
the Postdoc-Program of the German Academic Exchange
Service (DAAD), from the U.S. department of Energy
under grant DE-FG02-05ER41367 and within the framework of the Beam Energy Scan Theory (BEST) Topical Collaboration.
D.B., Iu.K. and H.P. thank the Institute for Nuclear Theory at the University of Washington for its hospitality and the Department of Energy for partial support during the completion of this work.
\end{acknowledgments}
% 
%______________________________________________________________________ 

\begin{appendix}
\section{Directed flow}
\label{app:A}
\begin{figure*}[!h]
\includegraphics[scale=0.9]{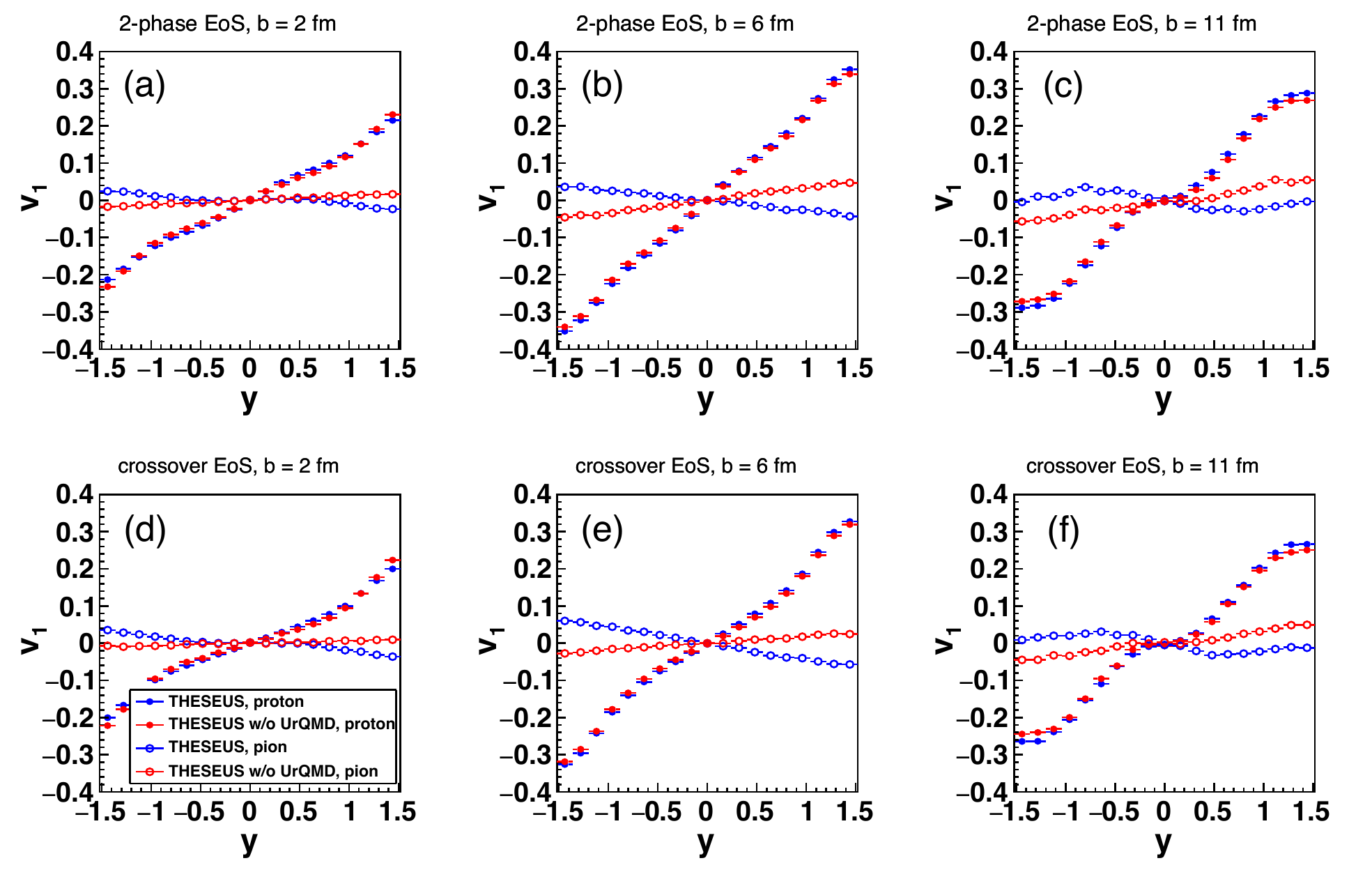}
\caption{(Color online)
Directed flow ($v_1$) of protons (full symbols) and pions (open symbols) for central ($b=2$ fm), semicentral ($b=6$ fm) and peripheral ($b=11$ fm) Au+Au collisions at $E_{\rm lab}=8~$A~GeV.  
The upper row (panels (a)-({c}) is for the 2-phase EoS while the lower row (panels (d)-(f)) shows 
results for the crossover EoS.
In each panel we show the direct comparison of THESEUS with (blue symbols) and without (red symbols)
UrQMD afterburner. Remarkable is the effect of turning pion flow to antiflow due to hadronic rescattering 
in the dense baryonic medium.
\label{8AGeV}
}
\end{figure*}

\begin{figure*}[!th]
\includegraphics[scale=0.9]{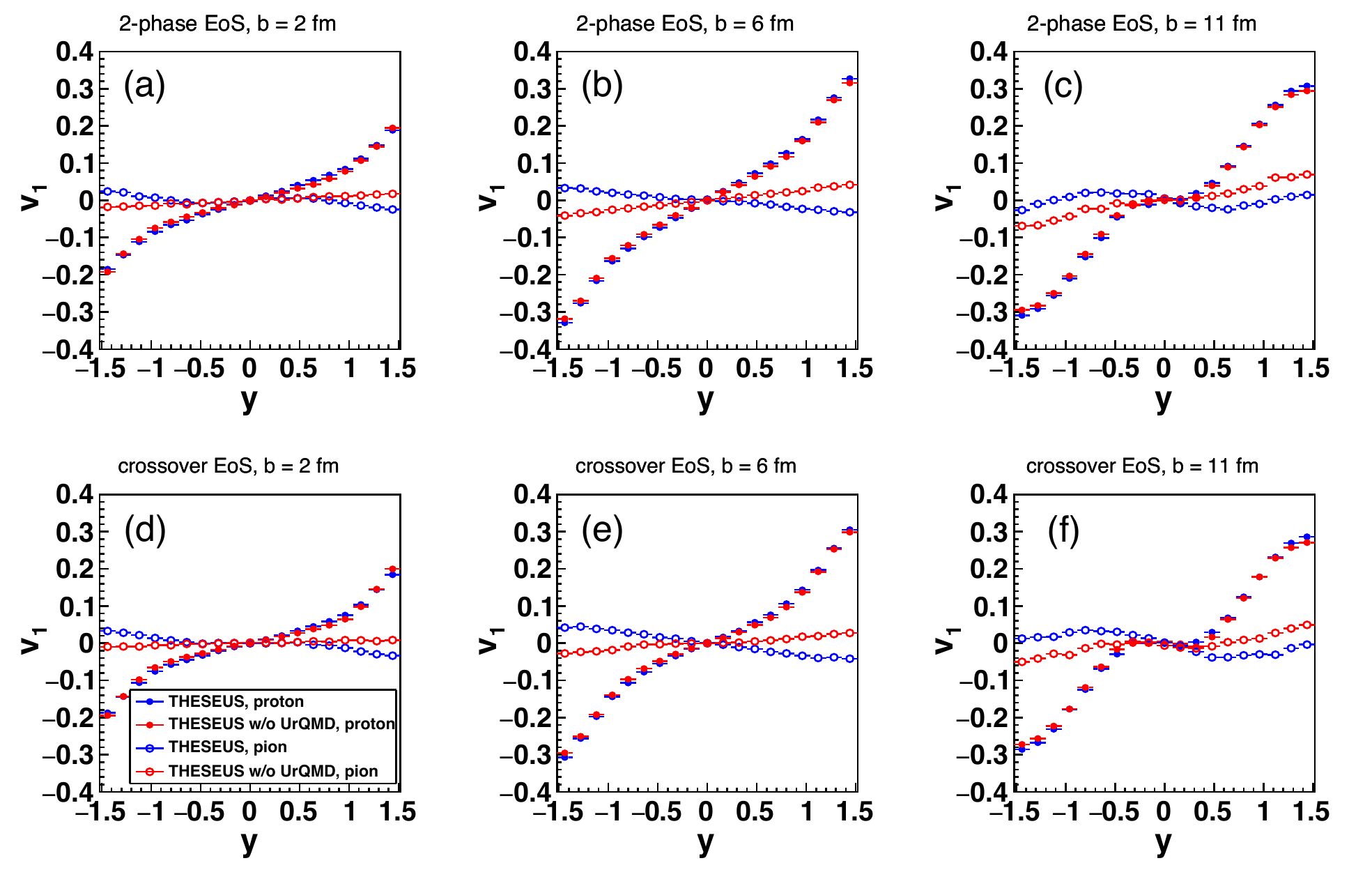}
\includegraphics[scale=0.9]{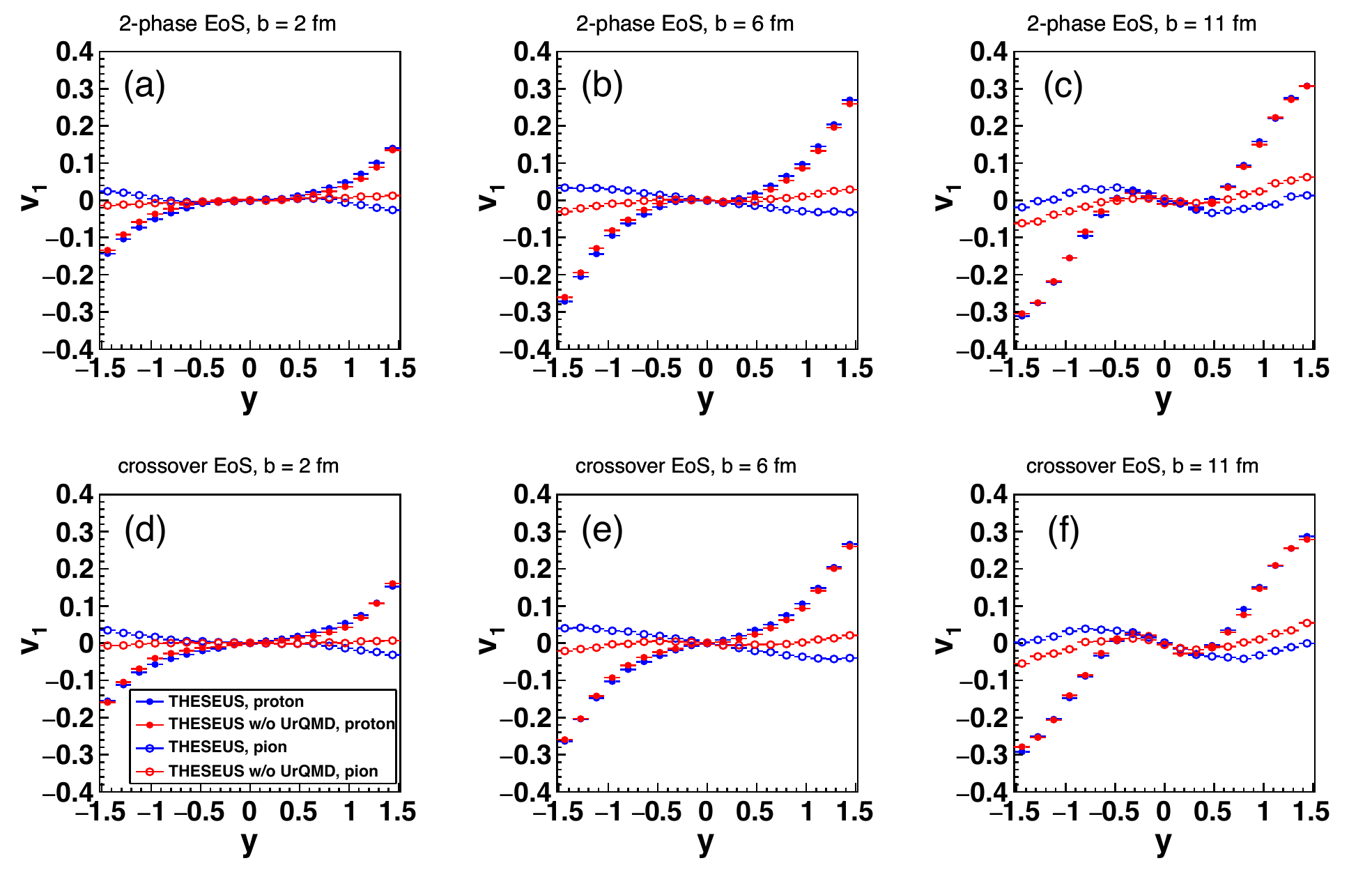}
\caption{(Color online)
Same as Fig.~\ref{8AGeV} for $E_{\rm lab}=10~$A~GeV (upper two rows) and $E_{\rm lab}=15~$A~GeV
(lower two rows).
\label{15AGeV}
}
\end{figure*}

\begin{figure*}[!th]
\includegraphics[scale=0.9]{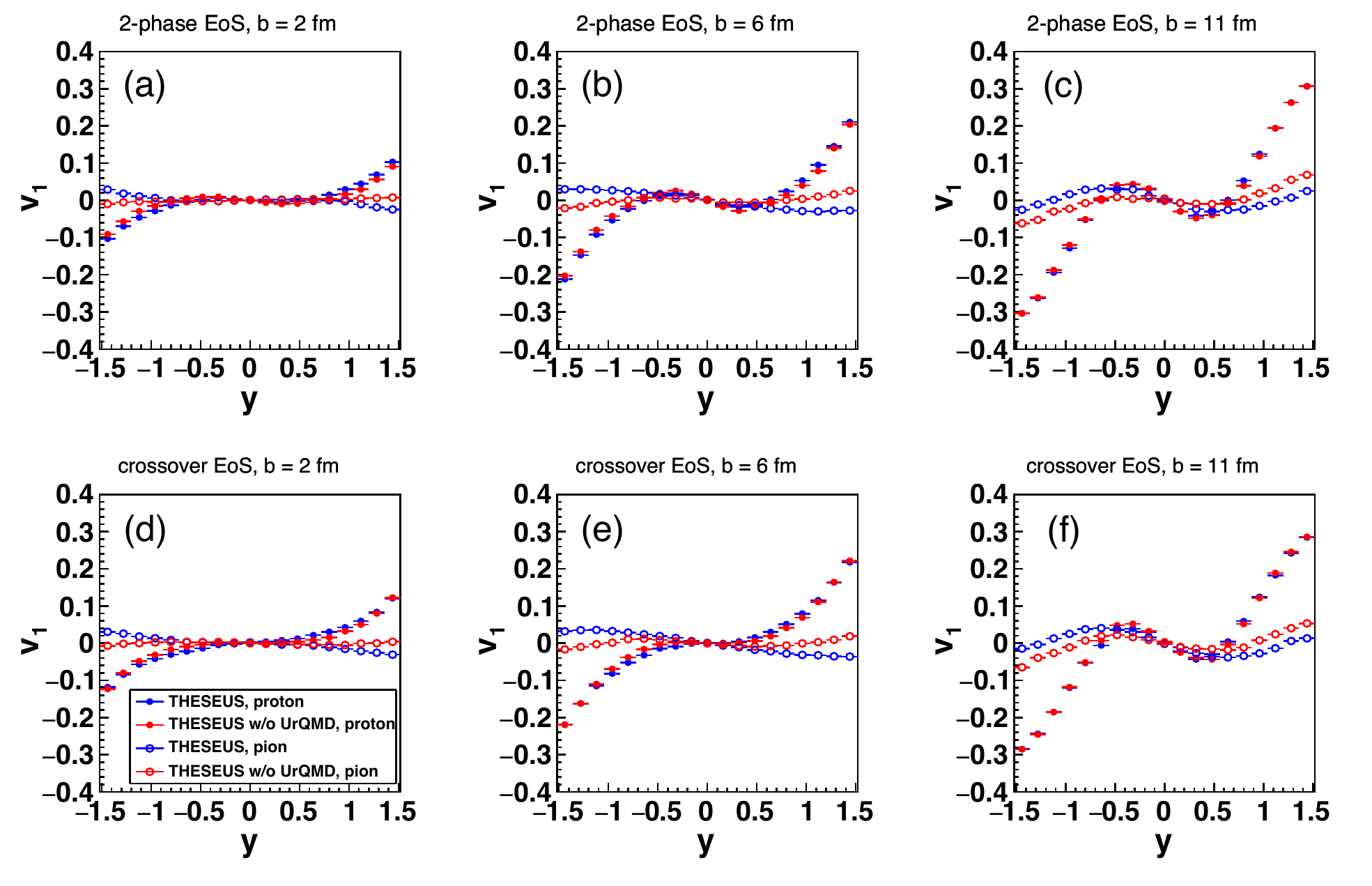}
\includegraphics[scale=0.9]{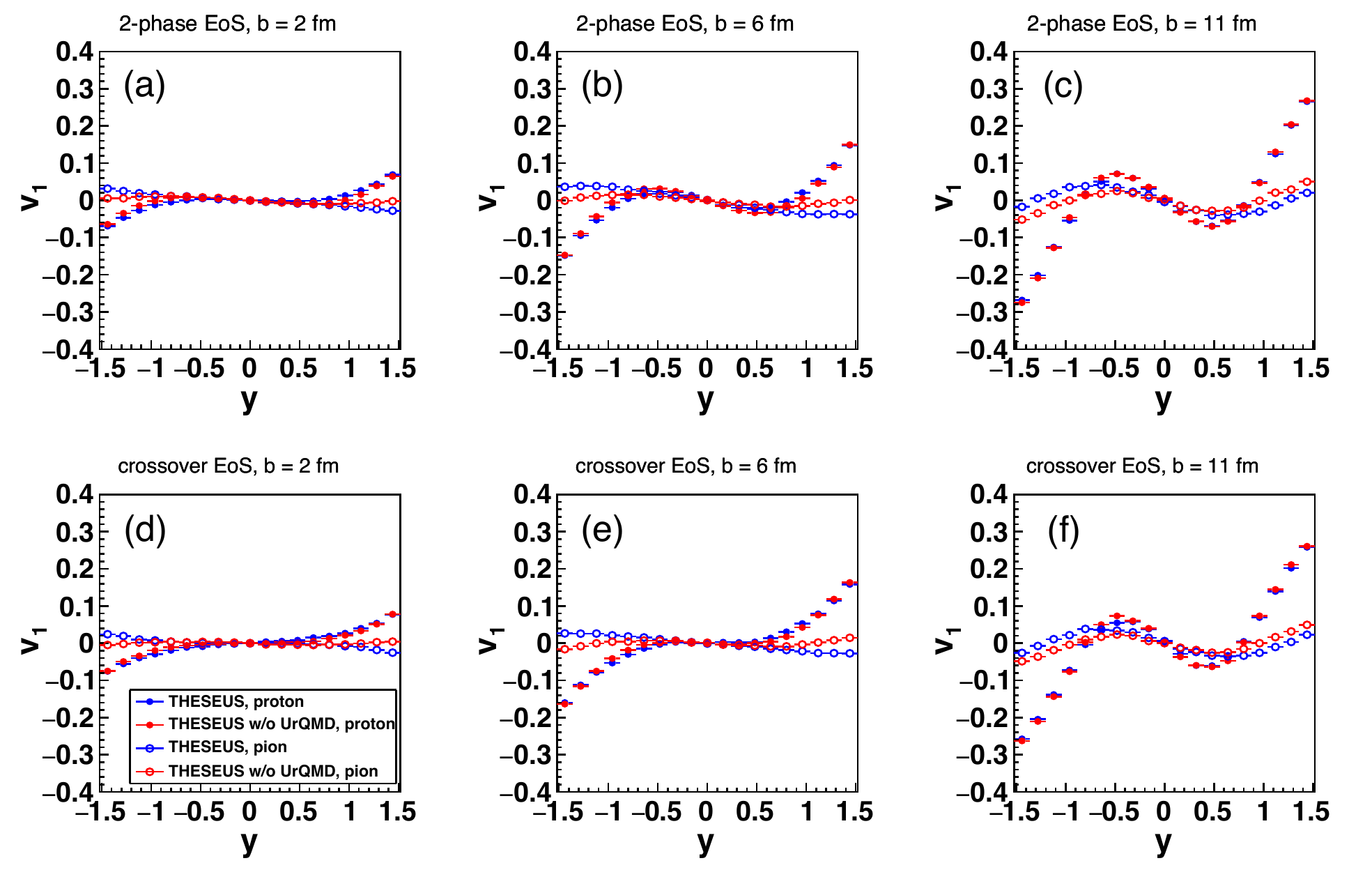}
\caption{(Color online)
Same as Fig.~\ref{8AGeV} for $E_{\rm lab}=20~$A~GeV (upper two rows) and $E_{\rm lab}=30~$A~GeV
(lower two rows).
\label{30AGeV}
}
\end{figure*}

\begin{figure*}[!th]
\includegraphics[scale=0.9]{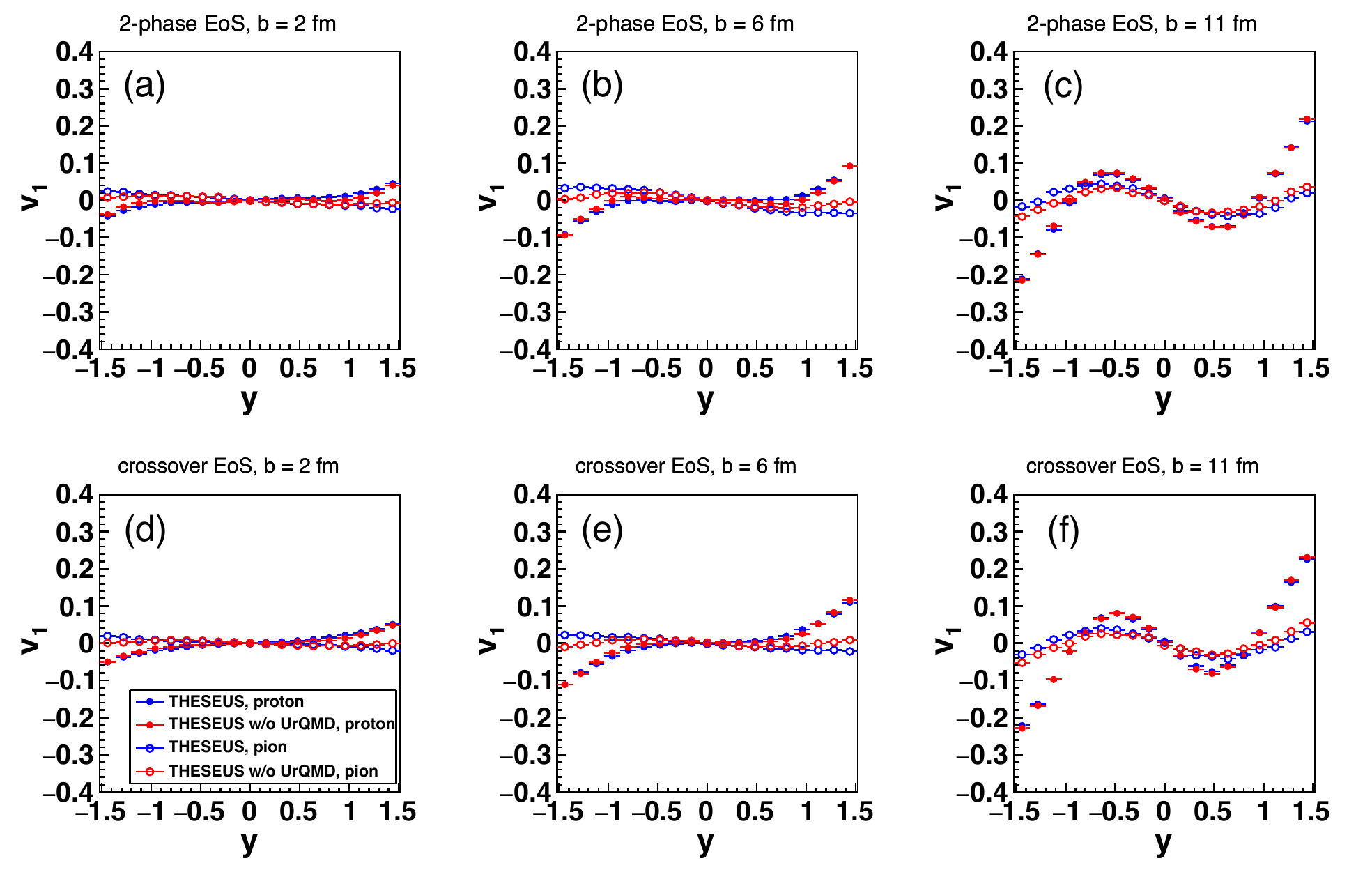}
\includegraphics[scale=0.9]{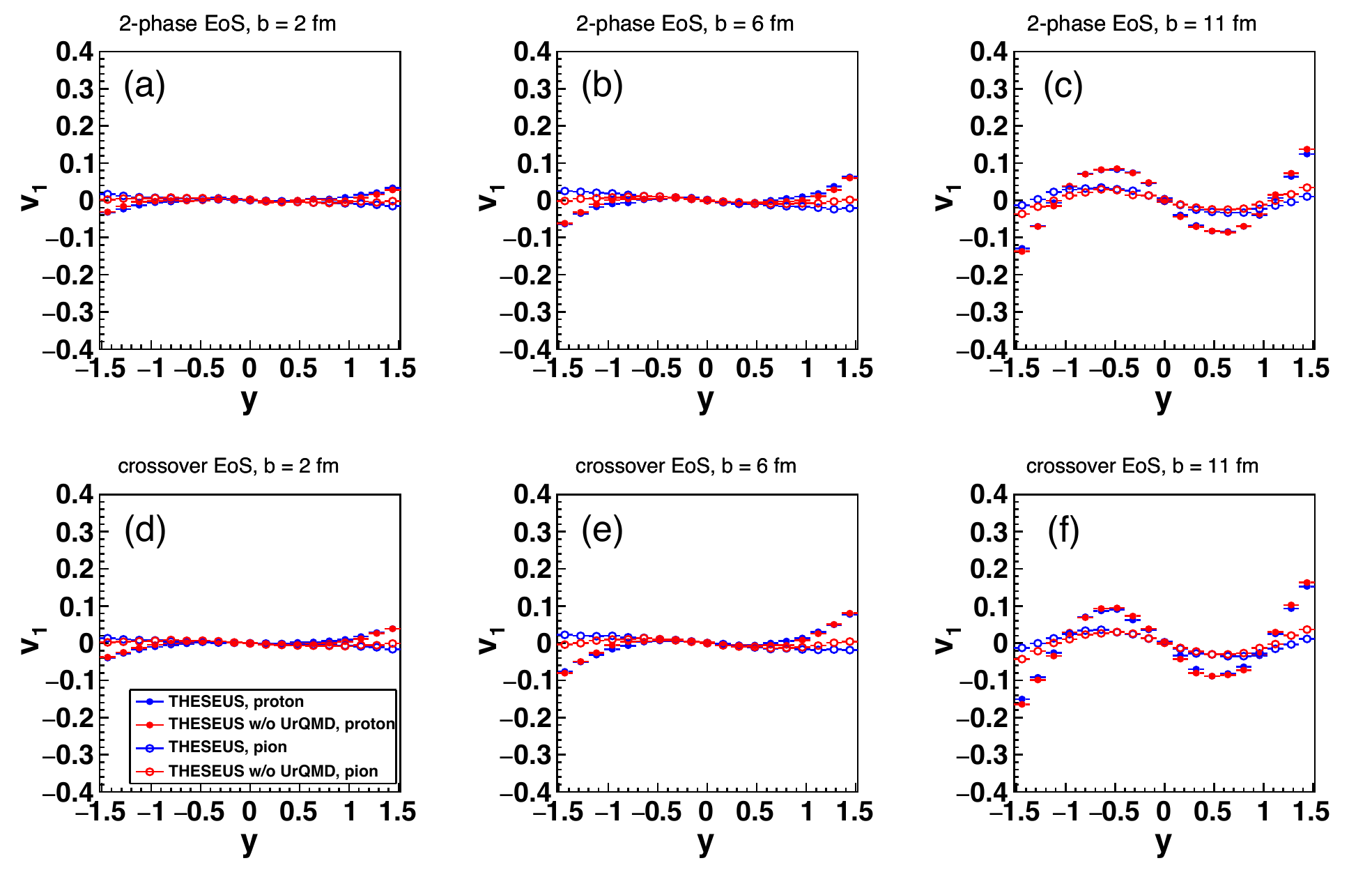}
\caption{(Color online)
Same as Fig.~\ref{8AGeV} for $E_{\rm lab}=43~$A~GeV (upper two rows) and $E_{\rm lab}=70~$A~GeV
(lower two rows).
\label{70AGeV}
}
\end{figure*}

In this Appendix we show the results of THESEUS with and without UrQMD afterburner for the directed flow 
$v_1$ of protons and pions at $E_{\rm lab}=8~$A~GeV (Fig.~\ref{8AGeV}), 10 and $15~$A~GeV 
(Fig.~\ref{15AGeV}), 20 and $30~$A~GeV (Fig.~\ref{30AGeV}), as well as 43 and $70~$A~GeV (Fig.~\ref{70AGeV}), 
comparing the case of the 2-phase EoS (first order phase transition, upper panels) with that of the crossover EoS (lower panels) at central (left panels), semicentral (middle panels) and peripheral (right panels) Au+Au collisions. 

These figures show the influence of hadronic final state interactions on the patterns of directed flow of protons and pions in the the broad rapidity range $-1.5 < y < 1.5$ and how it evolves from low energies 
in Fig.~\ref{8AGeV} to high energies in Fig.~\ref{70AGeV}.
At $E_{\rm lab}=8~$A~GeV in Fig.~\ref{8AGeV} we observe that hadronic rescattering causes the transition from flow to antiflow for pions due to the shadowing by a dense baryonic medium.
The flow of protons is not affected by the hadronic rescattering, which remains so for all energies.
The shadowing effect on the pion directed flow becomes less important at higher energies. 
At and above $30~$A~GeV hadronic rescattering has no effect on the directed flow of pions in the central rapidity region.

There is hardly any difference to be noticed in the pion directed flow patterns between the case of a 2-phase EoS and a crossover EoS.

The energy dependence of the slope of the proton directed flow exhibits a change of sign in the central rapidity region for semicentral collisions which is most pronounced at 20 - $30~$A~GeV for the 2-phase EoS. Since this pattern is absent for the crossover EoS it may be linked to the first order phase transition.
    
These features are displayed in a more compact manner in Figs.~\ref{fig-proton_v1_slope} and 
\ref{fig-pion_v1_slope} showing the energy scan of the slope of $v_1$ at midrapidity in semicentral Au+Au collisions for protons and pions, respectively.    

\section{Proton rapidity distribution}
\label{app:B}
\begin{figure*}[!h]
\includegraphics[scale=0.9]{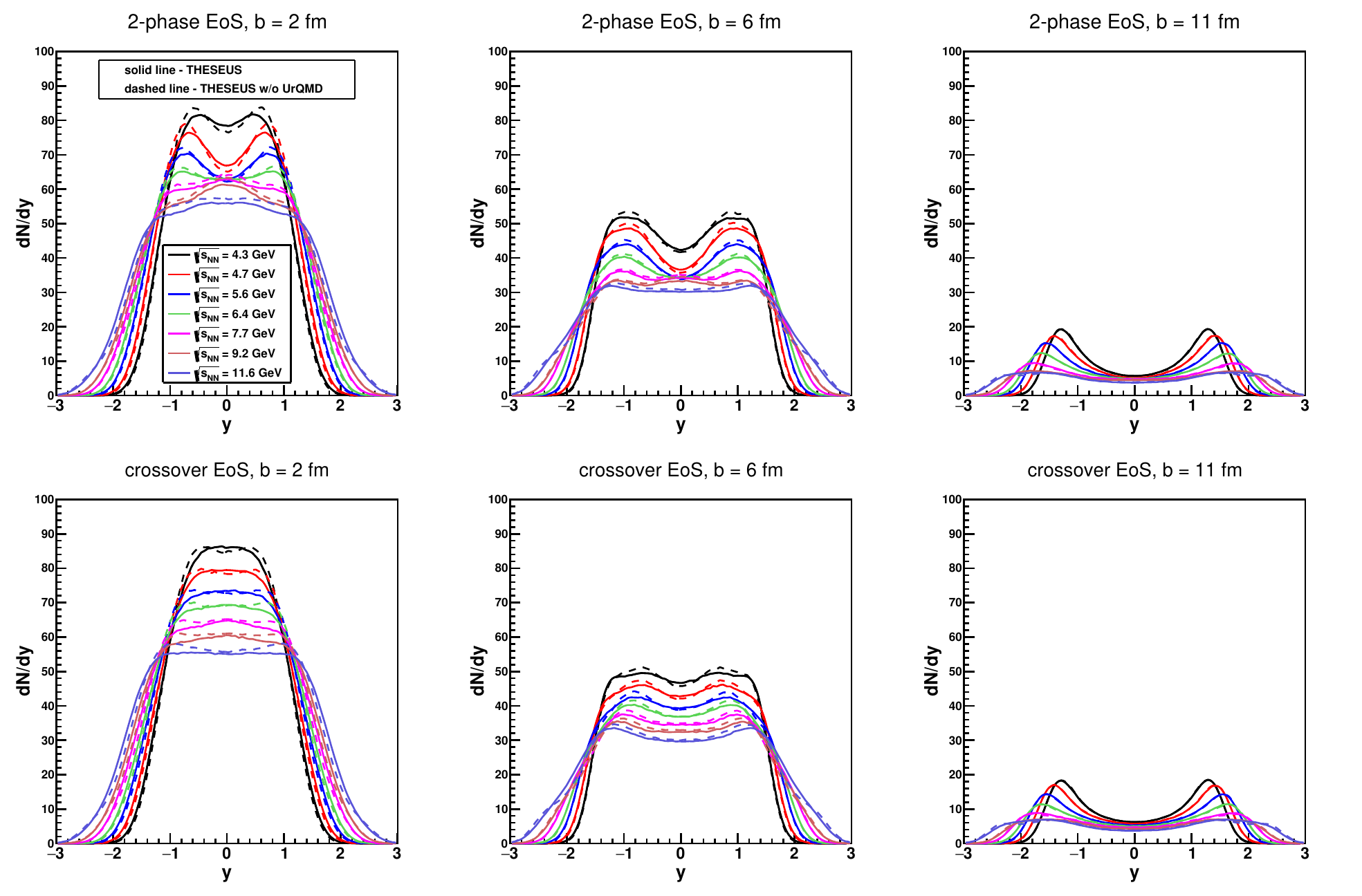}
\caption{(Color online)
Proton rapidity distributions for central ($b=2~$fm), semicentral ($b=6~$fm) and peripheral ($b=11~$fm) Au+Au collisions for the two-phase EoS (panels (a)-({c})) and for the crossover EoS (panels (d)-(f)). 
Each panel shows the results of THESEUS with (solid lines) and without (dashed lines) for the NICA energy scan with $\sqrt{s_{NN}}=4.3$, 4.7, 5.6, 6.4, 7.7, 9.3 and 11.6 GeV (different colors from black to light blue, resp.).
\label{p_rap}
}
\end{figure*}

In this Appendix we display the full proton rapidity distribution for seven energies of the NICA MPD energy scan (see Table~\ref{tab:energies}) in Fig.~\ref{p_rap} where for central collisions (left panels) a qualitative difference between the first order transition scenario of the 2-phase EoS (upper row) and the crossover transition scenario (lower row) can be observed. 
The hadronic final state interactions have only a minor effect on the "wiggle" in the energy scan of the 
curvature of the proton rapidity distribution at mid-rapidity as a signal of the first order phase transition,
see also Figs.~\ref{stopping_acceptance} and \ref{stopping_centrality} of Appendix \ref{app:C}.
For comparison the patterns at semicentral (middle panels) and peripheral collisions (right panels)
are also shown.  

\section{Influence of detector acceptance on the baryon stopping signal}
\label{app:C}
\begin{figure*}[!h]
\includegraphics[scale=0.92]{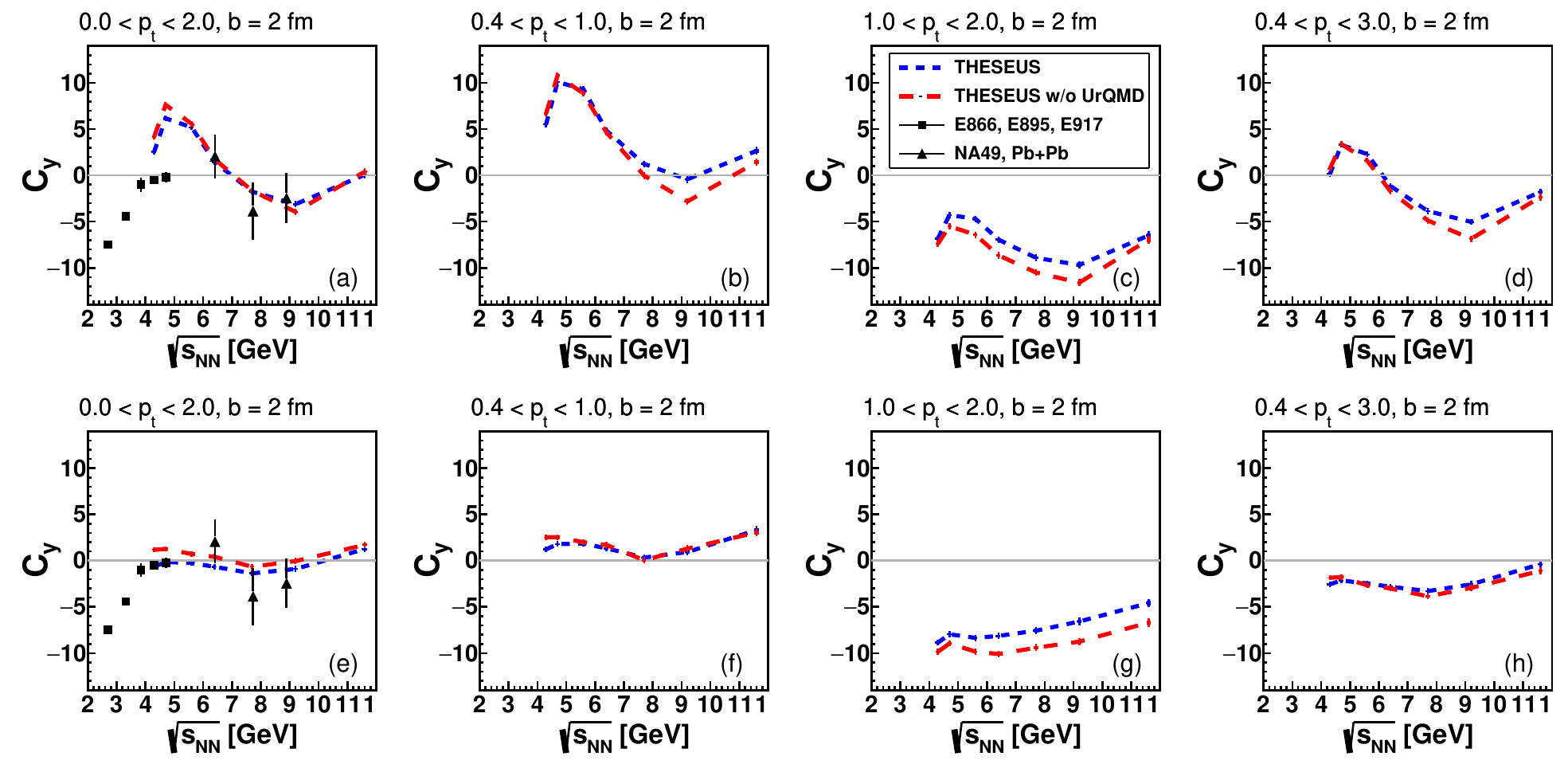}
\caption{(Color online)
Energy scan for the curvature $C_y$ of the net proton rapidity distribution at midrapidity for 
central Au+Au collisions with impact parameter $b=2~$fm in different acceptance windows for the 
transverse momentum $p_T$. 
We show results of THESEUS (blue short-dashed lines) and THESEUS without UrQMD (red long-dashed lines) together with presently available data (symbols in the panels (a) and (e)).
The results for the two-phase EoS (panels (a) - (d)) are compared to those for the crossover EoS (panels (e) - (h)).  
The "wiggle" as a characteristic feature for the EoS with a first order phase transition is rather robust against different acceptance cuts (see also Ref.~\cite{Ivanov:2015vna}) and hadronic final state interactions.
\label{stopping_acceptance}
} 
\end{figure*}

\begin{figure*}[!h]
\includegraphics[scale=0.8]{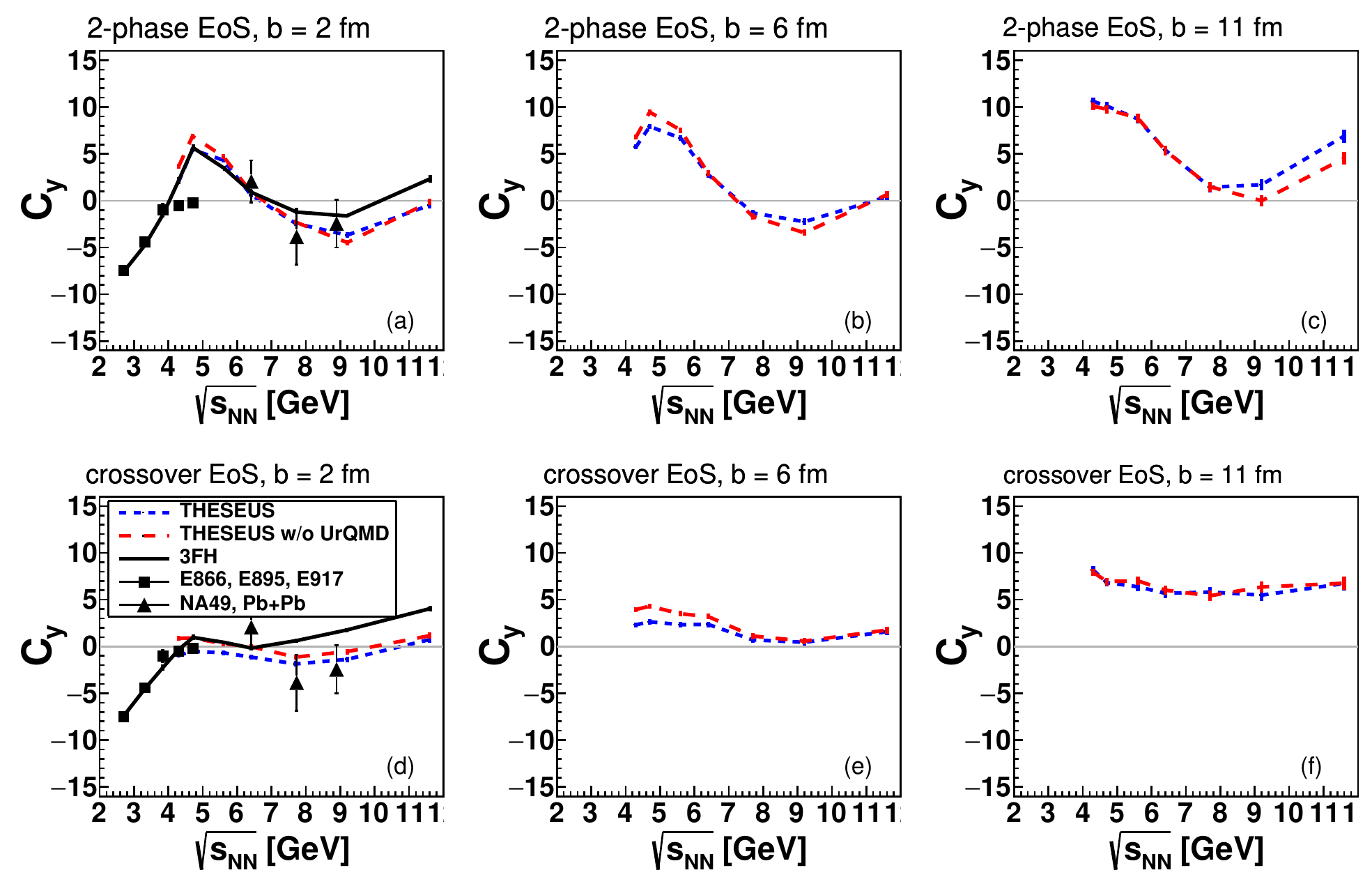}
\caption{(Color online)
Energy scan for the curvature $C_y$ of the net proton rapidity distribution at midrapidity for 
central Au+Au collisions with impact parameter $b=2~$fm (panels (a) and (d)), $b=6~$fm (panels (b) and (e)) and $b=11~$fm (panels ({c}) and (f)). 
We compare the 3FH model result (black solid lines) with THESEUS (blue short-dashed lines) and 
THESEUS without UrQMD (red long-dashed lines).
The results for the two-phase EoS (panels (a)-({c})) are compared to those for the crossover EoS 
(panels (d)-(f)).  
For noncentral collisions the curvature pattern is shifted towards positive values while the "wiggle" as a characteristic feature for the EoS with a first order phase transition remains rather robust.
\label{stopping_centrality}
} 
\end{figure*}

This Appendix is devoted to the illustration of the robustness of the baryon stopping signal for the 
first- order phase transition. 
For this purpose we show the response of the energy scan of the midrapidity curvature of the proton rapidity distribution to cuts in the proton transverse momentum spectrum in 
Fig.~\ref{stopping_acceptance} and to changes in centrality of the collision in Fig.~\ref{stopping_centrality}
for both cases, the 2-phase EoS (upper panels) and the crossover EoS (lower panels).
Comparing results of THESEUS (blue short-dashed lines) with those of THESEUS without final state interactions (red long-dashed lines) we confirm that the account for hadronic rescattering with the UrQMD afterburner has a minor, negligible effect which does not at all change the pattern:
The 2-phase EoS produces a "wiggle" structure while the crossover EoS results in a flat energy scan.
The comparison with the existing sparse data is so far not conclusive. 

From Fig.~\ref{stopping_acceptance} one learns that a restriction to high-$p_T$ events would reduce the
peak-dip difference of the wiggle structure of the 2-phase EoS and make it more similar to the flat 
pattern of the crossover EoS. 
The acceptance windows of the NICA MPD experiment \cite{Merz:2012} (second-to-left column) and   of
the STAR experiment at RHIC \cite{NuXu:2015} (rightmost column) are suitable to disentangle both cases. 
In the case of the STAR experiment, the upgrade to lower energies would be required to cover the whole
energy range of the wiggle.

From Fig.~\ref{stopping_centrality} we learn that triggering on central collisions is beneficial for the search for the wiggle signal of the first order phase transition. In this case the peak-dip structure includes a sign change of the midrapidity curvature of the proton rapidity disribution for the 2-phase EoS!
While the wiggle structure remains basically intact also for noncentral collisions, it gets shifted towards positive curvatures only which would make its identification less unambiguous.

\end{appendix}
\end{document}